\documentclass[fleqn,10pt]{wlscirep}
\usepackage[utf8]{inputenc}
\usepackage[T1]{fontenc}
\usepackage{graphicx}    
\usepackage{amsmath}    
\usepackage{amssymb}    
\newcommand{\mnras}{Mon. Not. R. Astron. Soc.}
\newcommand{\apj}{Astrophys. J.}
\newcommand{\apjl}{Astrophys. J.}
\newcommand{\apjs}{Astrophys. J.}
\newcommand{\aj}{Astron. J.}
\newcommand{\aap}{Astron. Astrophys.}

\newcommand{\pasp}{Publ. Astron. Soc. Pacific}

\newcommand{\kms}{\mbox{$\mathrm{km\,s^{-1}}$}}
\newcommand{\Ion}[2]{#1{\,\scriptsize #2}}
\newcommand{\teff}{\mbox{T$_{\rm eff}$}}
\newcommand{\logg}{\mbox{$\log g$}}

\title{Accurate mass and radius determinations of a cool subdwarf in
  an eclipsing binary}

\author[1,2,*]{Alberto Rebassa-Mansergas}
\author[3]{Steven G. Parsons}
\author[3,4]{Vikram S. Dhillon}
\author[5]{Juanjuan Ren}
\author[3]{Stuart P. Littlefair}
\author[6]{Thomas R. Marsh}
\author[1,2]{Santiago Torres}
\affil[1]{Departament de F\'{\i}sica, Universitat Polit\`{e}cnica de Catalunya, c/Esteve Terrades 5, 08860 Castelldefels, Spain}
\affil[2]{Institut d'Estudis Espacials de Catalunya, Ed. Nexus-201, c/Gran Capit\`a 2-4, 08034 Barcelona, Spain}
\affil[3]{Department of Physics \& Astronomy, University of Sheffield, Sheffield S3 7RH, UK}
\affil[4]{Instituto de Astrof\'{i}sica de Canarias, Via Lactea s/n, La Laguna, E-38205 Tenerife, Spain}
\affil[5]{National Astronomical Observatories, Chinese Academy of Sciences, 100012 Beijing, P. R. China}
\affil[6]{Department of Physics, Gibbet Hill Road, University of Warwick, Coventry, CV4 7AL, UK}
\affil[*]{alberto.rebassa@upc.edu}

\begin{abstract}
Cool subdwarfs  are metal-poor low-mass  stars that formed  during the
early  stages  of the  evolution  of  our  Galaxy.  Because  they  are
relatively  rare in  the vicinity  of  the Sun,  we know  of few  cool
subdwarfs in the solar neighbourhood, and  none with both the mass and
the radius  accurately determined.  This hampers  our understanding of
stars at  the low-mass end  of the  main-sequence. Here we  report the
discovery   of  SDSS\,J235524.29+044855.7   as  an   eclipsing  binary
containing a  cool subdwarf star,  with a white dwarf  companion. From
the  light-curve  and  the  radial-velocity curve  of  the  binary  we
determine the mass  and the radius of the cool  subdwarf and we derive
its  effective temperature  and luminosity  by analysing  its spectral
energy   distribution.    Our   results   validate   the   theoretical
mass-radius-effective  temperature-luminosity  relations for  low-mass
low-metallicity stars.
\end{abstract}
\begin{document}

\flushbottom
\maketitle
%
%
\thispagestyle{empty}


\section*{Introduction}

Cool  subdwarfs  are  metal-poor,   low-mass  main-sequence  stars  of
spectral type from mid K to late  M. They are generally referred to as
sdM/sdK stars.   The low metallicity  in cool subdwarfs  decreases the
opacity of their outer layers resulting  in a small radius as compared
to   a  main-sequence   star   of  the   same  effective   temperature
\cite{Burrows1993}. Thus, cool subdwarfs  have lower luminosities than
dwarfs  of the  same  effective  temperature and  lie  below the  main
sequence on the Hertzprung-Russell  diagram \cite{Morgan1943}. The low
metallicity of cool subdwarfs implies that they are typically very old
($\gtrapprox$  10  Gyr)  \cite{Burgasser2003}  and  are  part  of  old
Galactic populations  such as the thick  disc, the halo and  the bulge
\cite{Digby2003}.  Hence, cool subdwarfs formed in the first phases of
our  Galaxy and  therefore carry  important information  regarding its
structure  and  chemical evolution.   Moreover,  the  analysis of  the
spectral  energy distribution  (SED)  of cool  subdwarfs improves  our
understanding of the role of  metallicity in the opacity structure and
evolution  of cool  atmospheres  \cite{Rajpurohit2014}.  Depending  on
their  metallicity,  cool  subdwarf   stars  are  divided  into  three
sub-classes: sd  (subdwarfs), esd  (extreme subdwarfs) and  usd (ultra
subdwarfs) \cite{Lepine2007}, with  the metallicity content decreasing
from the sd to the usd stars.  Specifically, the iron abundance [Fe/H]
ranges  from $-0.34$  to $-0.87$  dex for  sd stars,  from $-0.87$  to
$-1.36$ dex  for esd stars  and it is lower  than $-1.36$ dex  for usd
stars \cite{Woolf2009}.  This classification is based  on the strength
of the TiO and CaH bands in the observed spectra \cite{Gizis1997}.

In order to understand the nature of these rare and important objects,
it is essential to determine  their most basic stellar parameters such
as masses  and radii.  To  that end,  and despite their  intrinsic low
luminosity, modern large-scale  surveys such as the  Sloan Digital Sky
Survey (SDSS) \cite{Stoughton2002} and the Large Sky Area Multi-Object
Fiber Spectroscopic Telescope  (LAMOST) \cite{Cui2012} have identified
$\sim$5000    cool   subdwarfs    during    the    last   few    years
\cite{Savcheva2014,  Bai2016}.   A recent  ground-based  spectroscopic
survey  allowed radii  determinations  of 88  cool  subdwarfs via  the
analysis   of  their   SEDs   and  the   use   of  $Gaia$   parallaxes
\cite{Kesseli2018}.   However,  we currently  know  of  only six  cool
subdwarfs   for   which   dynamical    masses   have   been   measured
\cite{Jao2016}. Moreover,  it has  to be stressed  that no  mass $and$
radius values for  a single cool subdwarf have been  measured to date.
Thus, the  bottom-end of the  main sequence remains  unconstrained for
metal-poor stars.

\begin{figure}[ht]
\centering
\includegraphics[angle=-90,width=0.6\linewidth]{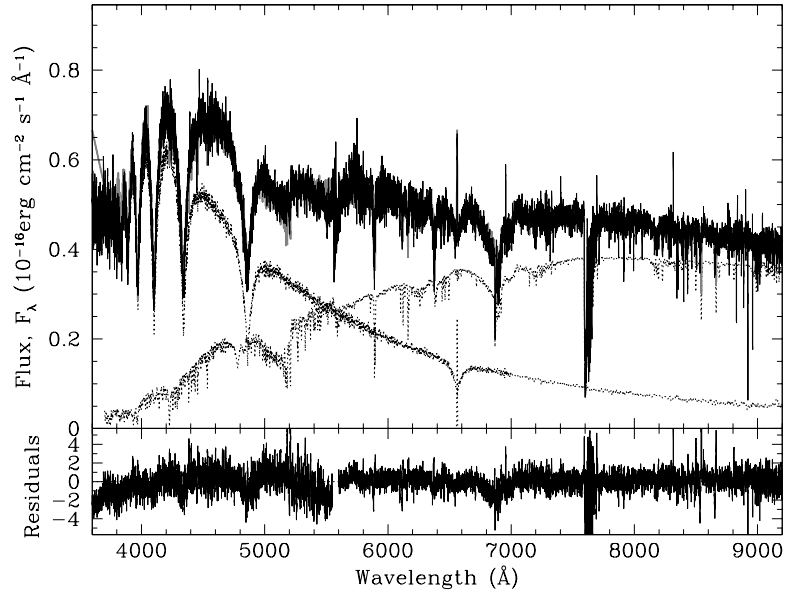}
\caption{Top  panel: X-Shooter  optical  spectrum of  SDSS\,J2355+0448
  (black solid  lines). The hydrogen  Balmer absorption lines of  a DA
  white dwarf can be clearly seen  in the blue.  At longer wavelengths
  the spectrum is mostly featureless,  with the exception of the broad
  CaH molecular band at $\sim$6800\AA,  typical of cool subdwarfs. The
  strong  H$\alpha$  emission  arises  most  likely  due  to  magnetic
  activity of  the cool subdwarf  and/or from wind accretion  onto the
  white  dwarf. The  two-component template  spectra (white  dwarf and
  cool  subdwarf) that  best fit  the observed  spectrum are  shown as
  black dotted  lines. The bottom  panel shows the residuals  from the
  fit. The spectrum has been binned by a factor of five for clarity.}
\label{fig:fitall}
\end{figure}

Here we report the  discovery of SDSS\,J235524.29+044855.7 (hereafter
SDSS\,J2355+0448) as  an eclipsing binary containing  a cool subdwarf,
with  a  white  dwarf   companion.   SDSS\,J2355+0448  was  originally
identified  as a  white  dwarf candidate  \cite{Kepler2015} and  later
re-discovered as  part of  our systematic search  of white  dwarf plus
low-mass   main  sequence   binaries  within   the  SDSS   and  LAMOST
spectroscopic data bases \cite{Rebassa2016,  Ren2018} (see the optical
spectrum in  Figure\,\ref{fig:fitall}).  Eclipsing binaries  offer the
opportunity  to measure  directly  the  masses and  radii  of the  two
components           with            unprecedented           precision
\cite{parsons2017,parsons2018}. This is the case for the cool subdwarf
in SDSS\,J2355+0448.

\section*{Results}

Available photometry  from the  Catalina Sky  Surveys \cite{Drake2009}
revealed  the binary  star SDSS\,J2355+0448  to be  eclipsing with  an
orbital period of 2.15 hours. We were awarded Director's Discretionary
Time  at the  8.2m Very  Large Telescope  to obtain  X-Shooter spectra
covering  the entire  optical  plus near-infrared  wavelengths of  the
target   (see  the   Methods  section   for  a   description  of   the
observations). From the  spectra we derived the  white dwarf effective
temperature and surface gravity (hence  its cooling age), the systemic
velocity of  the binary,  the radial-velocity  curve amplitude  of the
cool subdwarf  and its  metallicity class, and  we estimated  the cool
subdwarf's [Fe/H] abundance.

We  were also  awarded  time  on the  10.4m  Gran Telescopio  Canarias
equipped with HiPERCAM \cite{dhillon2018}  to sample the orbital light
curves in five filters (see also the Methods section for a description
of the observations). The data revealed  not only the primary but also
the secondary eclipse (i.e. the transit of the white dwarf in front of
the  cool subdwarf).  The  latter  is crucial  in  fixing the  orbital
inclination and stellar radii relative  to the orbital separation. The
analysis   of   the   HiPERCAM    light-curve,   together   with   the
radial-velocity   curve  of   the  cool   subdwarf,  allowed   precise
determinations of the  masses and radii of both  stellar components as
well   as  measurements   of  the   SDSS  magnitudes   for  the   cool
subdwarf.  These  magnitudes,  together with  available  near-infrared
photometry of the binary, permitted  the effective temperature and the
bolometric luminosity  of the cool  subdwarf to be determined  via SED
fitting.

The   fitted   and   derived   stellar  and   binary   parameters   of
SDSS\,J2355+0448 are provided in Table\,\ref{tab:params}.

\subsection*{The white dwarf stellar parameters}
\label{s:wdparam}

The   optical    spectrum   of   SDSS\,J2355+0448   (top    panel   of
Figure\,\ref{fig:fitall})  clearly  shows  the white  dwarf  and  cool
subdwarf components.   To derive the  stellar parameters of  the white
dwarf via  Balmer-line fitting  (namely the effective  temperature and
the surface  gravity) it  is necessary to  subtract the  cool subdwarf
contribution. To  that end we applied  a decomposition/fitting routine
\cite{Rebassa2007} to the individual  X-Shooter optical spectra of the
binary. Note that we did not use a combined optical X-Shooter spectrum
for  this  purpose since  we  were  not  able  to measure  the  radial
velocities  of the  white  dwarf,  and hence  the  Balmer  lines in  a
combined spectrum would have suffered from orbital smearing.

The  decomposition/fitting   routine  first  fits   the  two-composite
spectrum with  a grid of observed  cool subdwarf \cite{Rajpurohit2016}
and  white  dwarf  \cite{Rebassa2007}   templates  and  subtracts  the
best-fitted cool subdwarf template,  appropriately scaled in flux (see
the best two-composite fit to one  of our X-Shooter spectra in the top
panel  of Figure\,\ref{fig:fitall}).   In a  second step,  the routine
fits  both  the  normalised  Balmer   lines  and  the  whole  spectrum
(continuum plus  lines) of  the residual white  dwarf spectrum  with a
white  dwarf model  atmosphere grid  \cite{Koester2010} to  derive the
white   dwarf  effective   temperature   and   surface  gravity   (see
Supplementary    Figure\,1).     For    further    details    on    our
decomposition/fitting  routine  we point  the  reader  to the  Methods
section.

\begin{table}[ht]
\centering
\begin{tabular}{@{}lccccc@{}}
\hline
Parameter & units & value & uncertainty  \\
\hline
Orbital period & days & 0.089\,778\,006\,5 & 0.000\,000\,002\,4 \\
Binary inclination & degrees & 89.7 & 0.2 \\
Binary separation & R$_\odot$ & 0.711 & 0.003 \\
Mass ratio & M$_\mathrm{sd}$ / M$_\mathrm{WD}$ & 0.335 & 0.002 \\
Centre of WD eclipse & MJD(BTDB) & 58074.107\,683\,0 & 0.000\,008\,7 \\
$Gaia$ parallax & mas & 1.89 & 0.46 \\
Bayesian distance \cite{Bailer2018} & pc & 541 & -124 +216 \\
Right ascension & deg & 358.85133 & \\
Declination & deg & 4.81539 & \\
SDSS $g$ & mag & 19.59 & 0.01 \\
\hline
WD T$_\mathrm{eff}$ & K & 13,247 & 200 \\
WD $\log{g}$ & dex & 7.680 & 0.075 \\
WD cooling age & Gyr & 0.40 & 0.09 \\
WD mass & M$_\odot$ & 0.4477 & 0.0052 \\
\hline
K$_\mathrm{sd}$ & $\mathrm{km\,s^{-1}}$ & 296.5 & 3.3 \\
$\gamma_\mathrm{sd}$ & $\mathrm{km\,s^{-1}}$ & -117.7 & 2.7 \\
sd T$_\mathrm{eff}$ & K & 3,650 & 50\\
sd bolometric luminosity & L$_\mathrm{\odot}$& 4.3$\times10^{-3}$& 2.4 $\times10^{-3}$\\
sd spectral sub-type & & usdK7 & $\pm1$ subclass\\
sd [Fe/H] abundance & dex & -1.55 & 0.25\\
sd mass & M$_\odot$ & 0.1501 & 0.0017 \\
sd radius (towards WD) & R$_\odot$ & 0.1821 & 0.0007 \\
sd volume-averaged radius & R$_\odot$ & 0.1669 & 0.0007 \\
\hline
\end{tabular}
\caption{\label{tab:params}  Fitted  and  derived stellar  and  binary
  parameters for SDSS\,J2355+0448.  sd and WD stand  for cool subdwarf
  and white dwarf,  respectively. We also include  the right ascension
  and  declination of  the binary,  the SDSS  $g$-band magnitude,  the
  $Gaia$ parallax and the distance.}
\end{table}

In most cases, the best-fitted template  of the cool subdwarf was that
of an  sdK7 spectral type, although  the sdM1 and sdM3  templates also
provided good fits to some of  the spectra. We emphasise that, at this
stage, we are not aiming at  determining the spectral type of the cool
subdwarf,  but only  the  stellar  parameters of  the  white dwarf  by
subtracting  the  cool subdwarf  flux  contribution.  The white  dwarf
effective temperatures obtained in this way from our X-Shooter optical
individual spectra range from 12,882$\pm$29\,K to 13,335$\pm$94\,K and
the   surface  gravities   between  7.59$\pm$0.03   and  7.76$\pm$0.03
dex.  Taking   the  average   values  we  find   13,247$\pm$76\,K  and
7.68$\pm$0.03 dex, respectively.  It has to be noted  however that the
uncertainties obtained are underestimated  due to systematic errors in
flux calibration,  in the  normalization process  and/or in  the model
atmosphere grid  used. Hence, we take  more conservative uncertainties
of  200\,K and  0.075 dex  in  our spectroscopic  values of  effective
temperature and  surface gravity,  respectively. For  completeness, we
interpolated the averaged values  of effective temperature and surface
gravity  in  white  dwarf   cooling  sequences  for  low  metallicites
\cite{Althaus2017} to  derive a (model dependent)  spectroscopic white
dwarf mass of 0.457$\pm$0.015 M$_\mathrm{\odot}$  and a cooling age of
0.40  $\pm$  0.09  Gyr.  The  spectroscopic white  dwarf  mass  is  in
agreement with the  mass obtained from our  light-curve fitting method
(see Table\,\ref{tab:params}).

\subsection*{The radial-velocity curve of the cool subdwarf}
\label{s:rvs}

\begin{figure}[ht]
\centering
\includegraphics[angle=-90,width=0.6\linewidth]{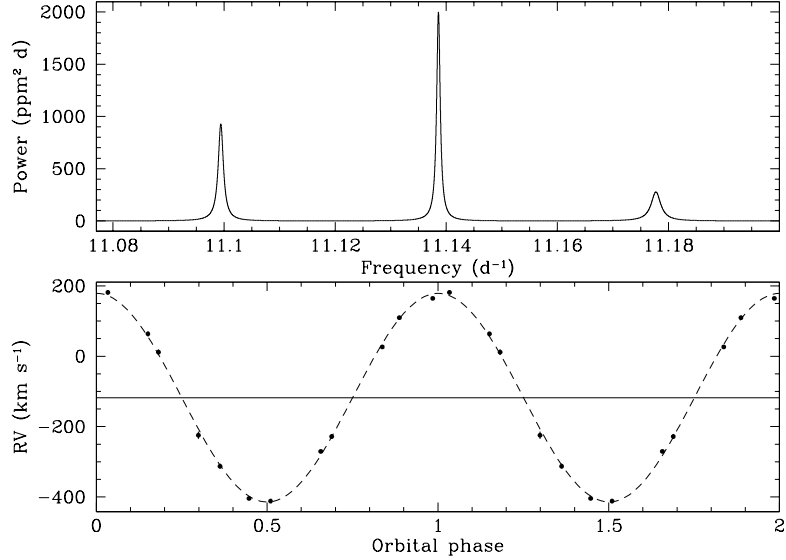}
\caption{Periodogram  obtained from  the radial-velocity  data of  the
  cool subdwarf in the binary SDSS\,J2355+0448, derived from X-Shooter
  spectroscopy. A clear  peak at 11.139\,d$^{-1}$ can  be seen. Bottom
  panel: the radial-velocity curve folded  over the period provided by
  the periodogram  in the top panel.  The radial-velocity $\pm1\sigma$
  error bars are shown but are too small to see.}
\label{fig:rv}
\end{figure}

We determined the radial velocities of  the cool subdwarf by fitting a
second-order  polynomial plus  a triple-Gaussian  line profile  to the
\Ion{Ca}{II}  absorption triplet  at  $\sim$8500\AA\,  sampled by  our
optical  X-Shooter  spectra  (see  the  Methods  section  for  a  full
description of the procedure). Periodograms calculated from the radial
velocities  to  investigate  the   periodic  nature  of  the  velocity
variations displayed  a clear  peak at 11.139\,d$^{-1}$  as well  as a
couple  of  weaker  aliases  due   to  the  sampling  pattern  of  the
observations  (Figure\,\ref{fig:rv},   top  panel).  We   carried  out
sine-fits of the form

\begin{equation}
\label{e-fit}
V_\mathrm{r} =
K_\mathrm{sd}\,\sin\left[\frac{2\pi(t-T_0)}{P_\mathrm{orb}}\right]
+\gamma
\end{equation}

\noindent
to  the  radial-velocity data  set,  where  $\gamma$ is  the  systemic
velocity, $K_\mathrm{sd}$ is the radial velocity semi-amplitude of the
cool subdwarf, $T_0$  is the time of inferior conjunction  of the cool
subdwarf, and  $P_\mathrm{orb}$ is  the orbital period.   The sine-fit
assumed the frequency corresponding to the strongest peak in the power
spectrum as  the orbital period  of the  binary, i.e. 2.154  hours, in
agreement  with the  orbital  period measured  from  the Catalina  Sky
Survey \cite{Drake2009} light curves. The radial-velocity curve folded
over  the  orbital  period  is  illustrated in  the  bottom  panel  of
Figure\,\ref{fig:rv}.

\subsection*{The mass and radius of the stellar components}
\label{s:massrad}

\begin{figure}[ht]
\centering
\includegraphics[width=0.8\linewidth]{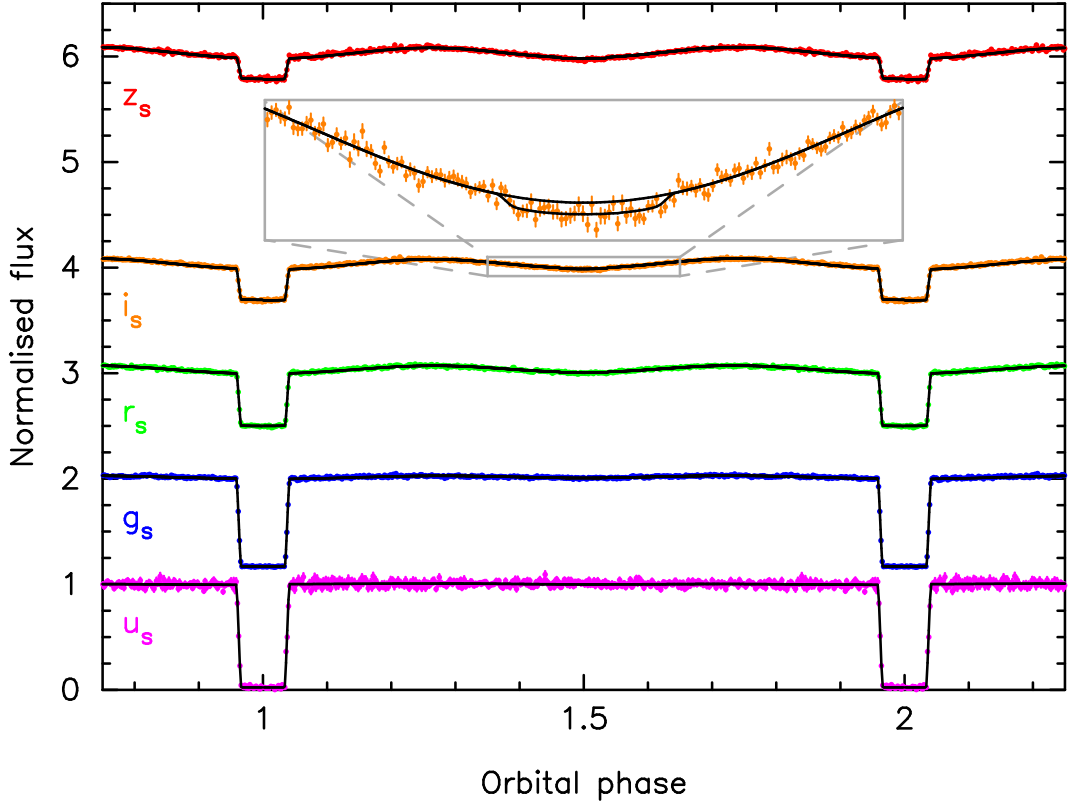}
\caption{Phase-folded HiPERCAM  light curves of  SDSS\,J2355+0448 with
  model fits over-plotted  (black lines). The $g_s$,  $r_s$, $i_s$ and
  $z_s$ band  data have been  binned by a  factor of three  and offset
  vertically  for  clarity.  We  show  a  zoom-in  to the  $i_s$  band
  light-curve, which displays the  secondary eclipse (i.e. the transit
  of the white  dwarf in front of the cool  subdwarf). Over-plotted is
  our best fit  model with the secondary eclipse turned  on and off to
  demonstrate  its  depth.  The   error  bars  represent  $\pm1\sigma$
  uncertainties.}
\label{fig:full_lc}
\end{figure}

The    high-speed    light     curves    of    SDSS\,2355+0448    (see
Figure~\ref{fig:full_lc}) revealed a sharp eclipse of the white dwarf,
with an ingress/egress lasting 80 seconds and a total eclipse duration
of 11  minutes.  As  was already apparent  from the  spectroscopy, the
white dwarf dominates  the overall flux in the $u_s$  and $g_s$ bands,
resulting in  a deep eclipse  and little out-of-eclipse  variation. At
longer wavelengths  the tidally distorted  shape of the  cool subdwarf
results in a double-peaked ellipsoidal modulation signal away from the
eclipse.   In  the   $i_s$  band  we  detect   the  secondary  eclipse
(Figure~\ref{fig:full_lc}).   All  of  these features  allowed  us  to
constrain  the  stellar and  binary  parameters  with minimal  use  of
theoretical models.

We    fitted   the    light    curves   using    the   {\sc    lcurve}
code\cite{copperwheat2010}, which is specifically designed for fitting
the light  curves of compact  binary systems (see the  Methods section
for  a detailed  description  of  the light  curve  model and  fitting
procedure).  Due  to   the  lack  of  a  direct   measurement  of  the
radial-velocity semi-amplitude  of the white  dwarf we were  unable to
determine the masses and radii of both stars completely independent of
theoretical models. We therefore forced  the radius of the white dwarf
to  follow a  theoretical  mass-radius relationship  when fitting  the
light  curve, hence  the  parameters  of the  cool  subdwarf are  only
dependent   upon    the   well   tested   white    dwarf   mass-radius
relationship\cite{parsons2017}.

\subsection*{The effective temperature and the bolometric luminosity of the cool subdwarf star}
\label{s:vosa}

The Virtual  Observatory SED Analyzer  (VOSA) \cite{Bayo08} is  a tool
that compares observed photometry,  gathered from a significant number
of compliant  VO catalogues,  to different collections  of theoretical
model  spectra   for  measuring  the  stellar   parameters  (effective
temperature, surface gravity, metallicity, bolometric luminosity) of a
given object/s. We  used VOSA to derive the  effective temperature and
bolometric  luminosity of  the  cool subdwarf  in SDSS\,J2355+0448.  A
description of  the methodology  employed by VOSA  is provided  in the
Methods section.

\begin{figure}[ht]
\centering
\includegraphics[width=0.6\linewidth]{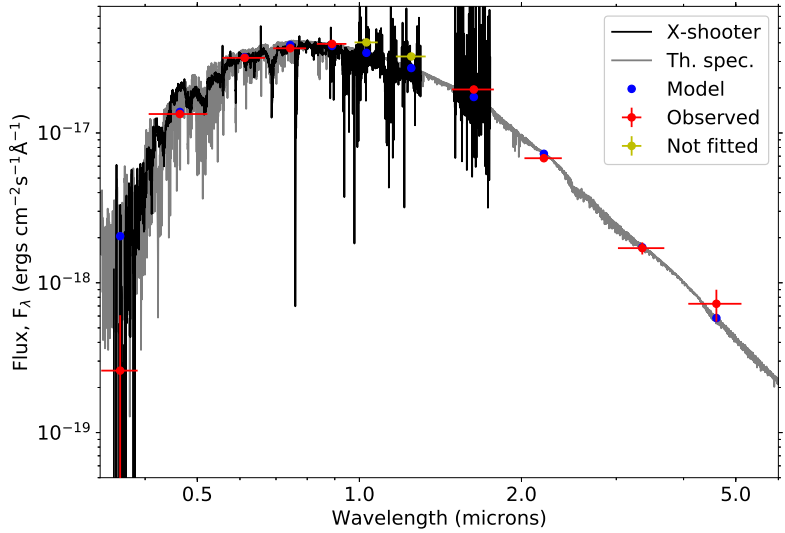}
\caption{The   observational    SED   of   the   cool    subdwarf   in
  SDSS\,J2355+0448 (black line, note  the white dwarf contribution has
  been  subtracted) and  its best-fit  model performed  by VOSA  (gray
  line).   The  observational  photometric  points are  shown  in  red
  (optical points  are obtained  by us directly  from the  light curve
  during  white dwarf  eclipse) and  the blue  dots are  the synthetic
  photometry. The yellow dots are UKIDSS $yj$ photometry that have not
  been used  in the  fit due  to white  dwarf flux  contamination. The
  error bars represent $\pm1\sigma$ uncertainties.}
\label{fig:vosa}
\end{figure}

Given  that  the white  dwarf  and  the  cool subdwarf  are  spatially
unresolved, we  did not make use  of VOSA to gather  available optical
photometry, since  those values  correspond to  the magnitudes  of the
binary system  and not  to the  cool subdwarf.  Hence, we  derived the
optical magnitudes of the cool subdwarf from our HiPERCAM light curves
during eclipse,  when the  flux contribution from  the white  dwarf is
zero.  The  HiPERCAM  magnitudes  were  converted  into  SDSS  $ugriz$
photometry. Colour terms between the HiPERCAM and the SDSS photometric
systems  were determined  by  folding  main-sequence template  spectra
\cite{Pickles1998}  through theoretical  band-passes for  HiPERCAM. We
also  considered  the  publicly available  near-infrared  $hk$  UKIDSS
\cite{Hewett06} magnitudes of SDSS\,J2355+0448,  where the white dwarf
flux contribution is negligible (the  $yj$ magnitudes do show signs of
contamination from the  flux of the white dwarf, as  revealed from the
X-Shooter   residual  cool   subdwarf  spectrum   that  results   from
subtracting   the   best-fit   white   dwarf   model   spectrum;   see
Fig\,\ref{fig:vosa})   and  the   far-infrared  WISE   \cite{Wright10}
$w_1w_2$ magnitudes, where  the white dwarf flux  contribution is also
negligible. For  each catalogue we adopted  a search radius of  3". No
2MASS  \cite{Skrutskie06},  VISTA  \cite{Cross12} or  extra  infra-red
magnitudes from other  surveys were found for our target.  In this way
we built the  observational SED of the cool subdwarf  from the optical
to the far-infrared wavelength range.

Two  physical parameters  were obtained  from the  VOSA SED  fits, the
effective  temperature and  the bolometric  luminosity. The  effective
temperature  was constrained  to  be 3600--3700\,K,  depending on  the
adopted model. Considering  that the intrinsic error  provided by VOSA
is  50\,K,  we assumed  a  value  of 3650  $\pm$  50\,K  for the  cool
subdwarf. The luminosities varied from (4.2 $\pm$ 2.4)$\times 10^{-3}$
L$_\mathrm{\odot}$    to     (4.4    $\pm$     2.4)$\times    10^{-3}$
L$_\mathrm{\odot}$  and  we  thus  adopted   a  value  of  (4.3  $\pm$
2.4)$\times   10^{-3}$   L$_\mathrm{\odot}$.  The   large   luminosity
uncertainty  is   directly  related  to  the   distance  error.  These
calculated values  of effective temperature and  bolometric luminosity
translate into a  radius of 0.164 $\pm$  0.050 R$_\mathrm{\odot}$ from
the  Stefan-Boltzmann  equation, in  good  agreement  with the  radius
derived   from   the   light-curve  fitting   (0.1669   $\pm$   0.0007
R$_\mathrm{\odot}$).

For  illustrative  purposes,  in Figure\,\ref{fig:vosa}  we  show  the
observational SED built together with the synthetic spectrum that best
fits  the photometric  data. The  synthetic spectrum  is an  excellent
match to the observed SED.

\subsection*{The metallicity class and the [Fe/H] abundance of the cool subdwarf star}

The parameter $\tau_\mathrm{TiO/CaH}$ \cite{Lepine2007, Lepine2013} is
a good indicator of the metallicity class of a dwarf star. It is based
on the  flux ratio  at specific wavelength  ranges and  quantifies the
weakening   of  the   TiO  band   strength  due   to  the   effect  of
metallicity. In order to calculate a  reliable value of this ratio for
the cool subdwarf, we subtracted the white dwarf contribution from the
X-Shooter spectrum  of SDSS\,J2355+0448. In  this case, since  we have
measured the radial  velocities of the cool subdwarf, we  were able to
correct the orbital  motion of the star and hence  combine all spectra
to one single (averaged) spectrum.  To subtract the white dwarf's flux
contribution we considered a white  dwarf model atmosphere spectrum of
the same effective  temperature and surface gravity  values as derived
for  the white  dwarf  in this  system (see  Table\,\ref{tab:params}),
obtained by  interpolating these  parameters in  our full  white dwarf
model atmosphere grid  \cite{Koester2010} and scaled in  flux. We thus
calculated a value of  $\tau_\mathrm{TiO/CaH}=0.095$ from the residual
cool  subdwarf spectrum,  i.e. an  ultra-low metallicity  usd spectral
class   \cite{Lepine2007,    Lepine2013}.   Using    three   different
calibrators,  this  translates  into  an iron  abundance  of  [Fe/H]$=
-1.6\pm0.2$     dex     \cite{Rajpurohit2014},    $-1.1\pm0.2$     dex
\cite{Mann2013}  and $-1.5\pm0.3$  dex \cite{Woolf2009}.  The possible
value for the  [Fe/H] abundance of the cool subdwarf  ranges then from
$-0.9$  to  $-1.8$  dex.  However, a  usd  metallicity  class  implies
[Fe/H]$<-1.35$  dex \cite{Woolf2009}  and we  hence assume  the [Fe/H]
abundance  should be  within  $-1.35$ and  $-1.8$  dex, i.e.  [Fe/H]$=
-1.55\pm0.25$ dex.  We also  made use  of iSpec  \cite{Blanco2014}, an
open source framework  for spectral analysis, to  attempt to determine
the [Fe/H]  abundance of  the cool subdwarf  star. However,  the value
obtained is subject  to large uncertainties ([Fe/H]$=-2.1  \pm 2$ dex)
due  to the  relatively  low signal  to  noise ratio  as  well as  the
intermediate resolution  of our spectra  (see the Methods  section for
details).

To determine the  spectral sub-type of the usd star  we re-applied our
decomposition/fitting  routine  to   the  averaged  X-Shooter  optical
spectrum,   using   a   usd   spectral   library   of   27   templates
\cite{Kesseli2018} instead. The spectral sub-type obtained in this way
for the low-metallicity star is usdK7.

\subsection*{The past and future evolution of SDSS\,J2355+0448}

Given its short  orbital period, the binary  star SDSS\,J2355+0448 has
very  likely evolved  in  the  past through  a  common envelope  phase
\cite{iben+livio1993}.   The  young cooling  age  of  the white  dwarf
implies such  an event took place  very recently and that  for most of
its life  the binary  star was  very likely  composed of  two low-mass
low-metallicity main-sequence  stars.  The  low mass measured  for the
white dwarf implies the common  envelope phase truncated the evolution
of  the  white  dwarf  precursor  when  it  was  ascending  the  giant
branch. Thus, the  white dwarf core is expected to  be composed mainly
of helium. It  has to be noted however that  the common envelope phase
is  not  expected  to  have   modified  the  parameters  of  the  cool
subdwarf. This is because secondary stars are not envisaged to be able
to accrete the  overflowing material during a  common envelope episode
due to their long  thermal time-scales \cite{Hurley2002}, a hypothesis
that  seems  to  be  observationally  confirmed  \cite{Parsons2018Md}.
Because the common envelope truncated the evolution of the white dwarf
progenitor, we  are not  able to derive  the white  dwarf's progenitor
mass  using  an  initial-to-final  mass relation  for  single  stellar
evolution \cite{Catalan2008, Cummings2018} nor  derive a total age for
the system. It is important to  emphasise however that the current age
of  SDSS\,J2355+0448 is  likely to  be very  old ($\gtrapprox$10  Gyr)
given the low metallicity of the  usd star. Indeed, the space velocity
with respect to the local standard of rest (U,V,W) $= (-36.5 \pm 29.3,
-208.3 \pm 24.8, 3.7 \pm 17.3)$ km/s derived from $Gaia$ data adopting
the systemic  velocity we measured  for the binary is  consistent with
halo membership for  this object, although we cannot rule  out a thick
disk origin \cite{Savcheva2014}. Due to  the short orbital period, the
cool subdwarf in SDSS\,J2355+0448 is tidally locked to the white dwarf
(the    ellipsoidal    modulation    can   clearly    be    seen    in
Figure\,\ref{fig:full_lc}),   therefore  its   rotational  period   is
commensurate  with the  orbital  period.  This  fast rotation  induces
chromospheric  activity  on  the  cool  subdwarf  (see  a  light-curve
displaying a flare in Supplementary Figure\,2).

We   used    the   binary    stellar   evolution    code   binary$\_$c
\cite{Izzard2004, Izzard2018}  (see a full description  in the Methods
section),  based on  the BSE  code \cite{Hurley2002},  to predict  the
future evolution  of SDSS\,J2355+0448.  The results  obtained indicate
that  a  phase of  stable  mass  transfer  will ensue  in  $\simeq$800
Myr.  The   system  will  then  become   a  semi-detached  cataclysmic
variable. We note however that the evolutionary timescale prior to the
onset of mass transfer should be  taken with caution since the angular
momentum loss mechanism/s that bring the  two stars closer are not yet
fully understood.

\begin{figure}[ht]
\centering
\includegraphics[angle=-90,width=0.9\linewidth]{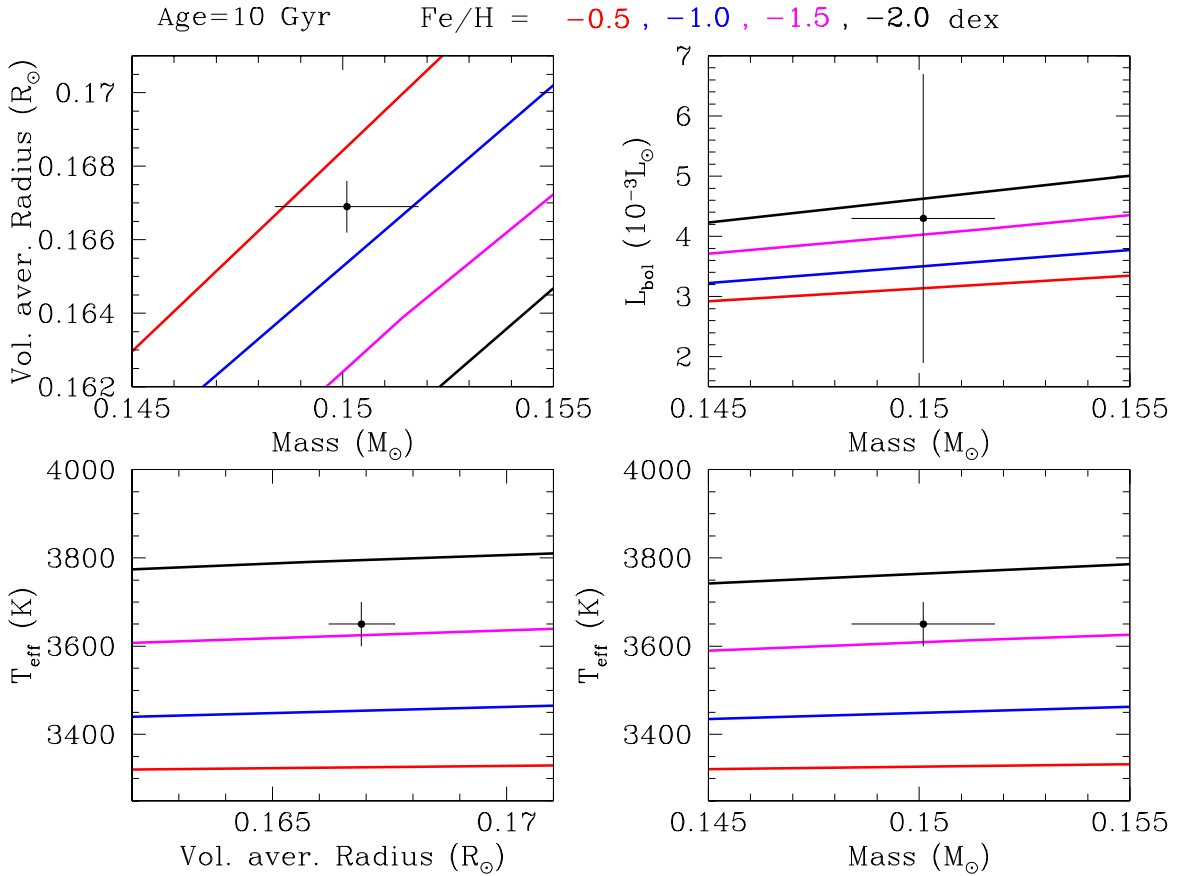}
\caption{The observed stellar parameter relations of the cool subdwarf
  in  SDSS\,J2355+0448 (black  solid  dots; the  error bars  represent
  $\pm1\sigma$ uncertainties) as compared  to the theoretical relations
  \cite{Dotter2008, Feiden2011} for an age of 10 Gyr and [Fe/H] values
  from -0.5 to -2.0 dex (solid lines).}
\label{fig:comp}
\end{figure}

\section*{Discussion}

We show a comparison between  the observed parameter relations for the
cool  subdwarf  in  SDSS\,J2355+0448   and  those  expected  from  the
theoretical  Dartmouth  isochrones  \cite{Dotter2008,  Feiden2011}  in
Figure\,\ref{fig:comp}.   Inspection of  the Figure  reveals that  the
theoretical  relations  agree  well  with  our  measured  values  when
assuming  an [Fe/H]  abundance  of $-1.5$  dex (a  value  which is  in
excellent agreement  with our estimate of  $-1.55\pm0.25$ dex), except
for  the theoretical  mass-radius  relation, which  yields a  slightly
overestimated  radius for  the measured  mass.  This  could be  due to
uncertainties in the adopted  limb-darkening coefficients for the cool
subdwarf when performing our light-curve  fit (see the Methods section
for further details).  Alternatively, an  over-sized radius could be a
consequence of the  cool subdwarf star being  magnetically active (see
Supplementary Figure\,2).   Over-inflated radii are often  measured for
magnetically active low-mass main-sequence  stars of solar metallicity
\cite{Parsons2018Md}.  It is also important  to note that the measured
radius  of  the   cool  subdwarf  is  undersized   compared  to  solar
metallicity models\cite{Baraffe1997, Feiden2011},  which is consistent
with the  theory that cool subdwarfs  should have smaller radii  for a
given effective temperature (see Supplementary Figure\,3).

In  this paper  we have  studied in  detail the  cool subdwarf  in the
eclipsing  binary SDSS\,J2355+0448.   The precise  mass and  radius we
have determined for this star  together with the effective temperature
and  luminosity we  have calculated  have allowed  us to  validate the
theoretical models of low-mass low-metallicity stars.

\section*{Methods}

\subsection*{Observations and their reduction}
\label{m:obs}

SDSS\,J2355+0448  was  observed  with the  medium  resolution  echelle
spectrograph  X-Shooter\cite{vernet2011}   on  the  8.2m   Very  Large
Telescope (VLT) in  Chile on the nights  of the 12th of  June 2018 and
the 8th of July 2018. X-Shooter  consists of three detectors (UVB, VIS
and  NIR)  which  obtain  simultaneous spectra  from  the  ultraviolet
atmospheric  cutoff   ($\sim$3000{\AA})  to  the  K   band  ($\sim$2.5
microns).  SDSS\,J2355+0448  was observed for  a full binary  orbit on
both nights, recording  a total of 12  UVB and VIS spectra  and 28 NIR
spectra. Exposure  times were  kept short (1073s,  1088s, 480s  in the
UVB, VIS and NIR arms respectively)  to minimise the amount of orbital
smearing  of the  lines. Due  to the  faintness of  the binary  in the
near-infrared   ($J=17.5$)  the   NIR  arm   spectra  have   very  low
signal-to-noise and were not used in any subsequent analysis. The data
were  reduced using  the  latest release  of  the X-Shooter  reduction
pipeline  (version  2.9.3).  Standard  reduction  steps  were used  to
debias,  flat-field, extract  and wavelength  calibrate the  data. The
spectra were  flux calibrated using spectra  of the spectrophotometric
standard  star Feige  110  obtained in  twilight at  the  end of  both
nights. Finally, all spectra were  placed on a heliocentric wavelength
scale.

High-speed  multi-band  photometry  of SDSS\,J2355+0448  was  obtained
using HiPERCAM\cite{dhillon2018} mounted on  the 10.4m Gran Telescopio
Canarias  (GTC). HiPERCAM  is  a quintuple-beam  imager equipped  with
frame-transfer CCDs allowing  simultaneous $u$, $g$, $r$,  $i$ and $z$
band imaging  at frames rates  of up to  1000 frames per  second. Note
that HiPERCAM  uses modified  versions of the  SDSS filters  with much
higher throughput, known  as Super-SDSS filters. These  are denoted as
$u_s  g_s r_s  i_s z_s$  to distinguish  them from  the standard  SDSS
filters ($ugriz$). Exposure times of 5s were used in the $g_s$, $r_s$,
$i_s$ and $z_s$ bands and 15s  in the $u_s$ band. SDSS\,J2355+0448 was
observed for a full orbit on the night of the 5th of October 2018, and
50 minutes of  data were obtained around the secondary  eclipse on the
nights of the 31st of October 2018  and 1st of November 2018. The data
were  reduced   using  the  dedicated  HiPERCAM   pipeline,  including
debiasing,   flat-fielding  and   fringe  correction   in  the   $z_s$
band.  Differential photometry  was  performed using  the nearby  star
SDSS\,J235517.21+045057.1  as  a  reference  source.  All  times  were
converted  to the  barycentric dynamical  timescale, corrected  to the
solar system barycentre, MJD(BTDB).

\subsection*{The spectral decomposition/fitting technique}
\label{m:decom}

The decomposition/fitting routine used in this work \cite{Rebassa2007}
was  developed to  derive  spectroscopic stellar  parameters of  white
dwarf plus  M dwarf binaries,  namely the  spectral sub-type of  the M
dwarf    and   its    radius,    and    the   effective    temperature
(T$_\mathrm{eff}$), surface gravity ($\log g$), mass and radius of the
white dwarf.   It uses an evolution  strategy \cite{rechenberg1994} to
decompose the  observed spectrum into  the two stellar  components. To
that end it optimises a  fitness function (i.e.  a weighted $\chi^2$).
The only difference between the  original routine and the one employed
in this  work is adopting a  series of cool subdwarf  template spectra
\cite{Rajpurohit2016, Kesseli2018} instead of  the original 10 M dwarf
templates.

The routine  fits first the  entire spectrum and selects  the best-fit
cool  subdwarf  template, which  is  scaled  in flux  and  subtracted.
Consequently, the residual white dwarf  spectrum is fitted with a grid
of    hydrogen-rich    white    dwarf   model    atmosphere    spectra
\cite{Koester2010}.   The grid  includes  1,339  spectra covering  the
6,000-100,000\,K T$_\mathrm{eff}$ and the 6.5-9.5 dex $\log g$ values.
The fit  is performed to the  normalised H$\beta$ to H10  Balmer lines
(the H$\alpha$  line is  excluded in  this process  since it  covers a
wavelength  range  of largest  residual  contamination  from the  cool
subdwarf). A bi-cubic  spline interpolation to the  $\chi^2$ values on
the  T$_\mathrm{eff}-\log  g$ grid  is  used  to derive  the  best-fit
T$_\mathrm{eff}$ and  $\log g$.  The errors  (1$\sigma$ uncertainties)
are  calculated projecting  the  $\Delta\chi^2=1$  contour around  the
$\chi^2$  of the  best  fit  into the  T$_\mathrm{eff}$  and $\log  g$
axes. This  results in a range  of parameters which are  averaged into
symmetric error bars (see the top panels of Supplementary Figure\,1).

Since the equivalent widths of the  Balmer lines of white dwarfs reach
a maximum for effective temperatures  near 13\,000\,K, the Balmer line
profile fits provide two sets of T$_\mathrm{eff}$ and $\log g$ values,
which are generally  known as the ``hot'' and  ``cold'' solutions.  In
other words, similar fits can be achieved on both temperature sides at
which  the  maximum  equivalent  width takes  place.   To  break  this
degeneracy the routine makes use of the model spectra grid for fitting
the entire  spectrum over the wavelength  range $3850-7150$\,\AA\ (see
the  bottom panel  of Supplementary  Figure\,1).  The  red part  of the
spectrum is not  considered by the fit due to  possible distortions in
the residual white dwarf spectrum arising from subtraction of the cool
subdwarf.  The T$_\mathrm{eff}$  and $\log g$ values  derived from the
fit  to   the  whole  spectrum  are   used  to  discard  one   of  the
``hot''/``cold''  solutions.  In  the  case  of SDSS\,J2355+0448,  the
best-fit T$_\mathrm{eff}$ and $\log g$  from the whole spectrum agrees
with the one obtained by the ``hot'' solution (see the top-right panel
of Supplementary Figure\,1).

\subsection*{The radial-velocity fitting method and the periodogram}
\label{m:rv}

We fitted  the Ca$_\mathrm{II}$ absorption triplet  at $\sim$8500\AA\,
sampled  by  the X-Shooter  VIS  arm  spectra with  a  triple-Gaussian
absorption profile of fixed separation (see an example in Supplementary
Figure\,4). To that  end we fixed the width of  each Gaussian, however
we  set the  three  amplitudes and  the radial  velocity  of the  cool
subdwarf star as free parameters.   This fitting method has been shown
to  provide both  accurate and  precise M  dwarf radial  velocities in
binary systems with white dwarf companions  such as the one studied in
this work \cite{Rebassa2017}.

We  run a  Lomb-Scargle periodogram  \cite{Scargle1982} to  the radial
velocities   of  the   cool   subdwarf   in  SDSS\,J2355+0448,   which
unfortunately  contained  several  aliases  as a  consequence  of  the
sampling   pattern   of  the   observations.    We   thus  tried   the
\textsf{ORT/TSA}\cite{Schwarzenberg-czerny1996}       command       in
\textsf{MIDAS},  which provided  a periodogram  with a  clear pick  at
11.139\,d$^{-1}$. The  \textsf{ORT/TSA} routine  uses a grid  of trial
periods  to fold  and phase-bin  the  data and  subsequently fits  the
folded radial velocity curve with a series of Fourier terms.

\subsection*{The light-curve fitting method}
\label{m:lc}

We  fitted   the  HiPERCAM  light  curves   of  SDSS\,J2355+0448  (see
Figure\,\ref{fig:full_lc})   using  a   code   written  for   binaries
containing  white  dwarfs \cite{copperwheat2010,  Bloemen2011}.   Each
stellar component in  the binary is subdivided into  small elements by
the program. This  is performed by following a geometry  that is fixed
by the radius  of the considered star as measured  along the direction
of centres towards  the other star. The code takes  into account limb-
and  gravity-darkening,  gravitational  microlensing,  Roche  geometry
distortion,  irradiation effects  and eclipses.  It uses  a non-linear
four-parameter model to implement limb-darkening \cite{claret2000}, of
the form
\begin{equation}
\frac{I(\mu)}{I(1)} = 1 - \displaystyle\sum_{k=1}^{4}a_{k}(1-\mu^{\frac{k}{2}}),
\end{equation}
being  $\mu =  \cos{\phi}$  (where  $\phi$ is  the  angle between  the
emergent flux  and the  line of sight),  and $I(1)$  the monochromatic
specific intensity  at the centre  of the stellar  disk.  Supplementary
Table\,1 lists  the adopted limb-darkening coefficients  for the white
dwarf\cite{gianninas2013} and cool subdwarf \cite{claret2011}.

The  model requires  a set  of  additional parameters  to be  defined,
namely    the    binary    inclination    $i$,    the    mass    ratio
$q=M_\mathrm{sd}/M_\mathrm{WD}$, the  orbital period $P_\mathrm{orb}$,
the    sum    of    the    unprojected    stellar    orbital    speeds
$V_s=(K_\mathrm{WD}+K_\mathrm{sd})/\sin{i}$,  the  mid-time  when  the
white dwarf eclipse occurs $T_0$, the radii of the stars scaled by the
orbital  separation $R_\mathrm{WD}/a$  and $R_\mathrm{sd}/a$,  and the
temperatures of the  stars T$_\mathrm{eff,WD}$ and T$_\mathrm{eff,sd}$
(is has to  emphasised that temperatures are  flux scaling parameters,
therefore correspond only approximately to the real temperatures). All
these parameters  were considered  as free  when fitting  the HiPERCAM
light curves,  the only exceptions  were the white  dwarf temperature,
which was fixed at 13,250\,K (Table\,\ref{tab:params}) and the orbital
period,  which was  fixed to  the value  found from  the Catalina  Sky
Survey results.  We  phase-folded all of our  HiPERCAM observations in
order to combine data taken on different nights.

The information provided by the profile  of the white dwarf eclipse is
not sufficient  to determine the  radii of both  stars as well  as the
orbital inclination.   As a consequence, the  fitting program requires
additional  information  \cite{parsons2017}  to break  the  degeneracy
between the scaled radii and the inclination.  This can be provided by
the distorted shape of the cool subdwarf owing to its proximity to the
white  dwarf,  which  results  in   a  double-peaked  feature  in  the
light-curve known  as ellipsoidal modulation. Unfortunately,  the fits
to  this  double-peaked  modulation  highly  depend  on  the  adopted
limb-darkening  coefficients for  the  cool subdwarf  as  well as  the
presence of  any star-spots. This  is a particularly large  problem in
the case of the cool subdwarf  star in SDSS\,J2355+0448 because of the
uncertainty in its metallicity. Adopting the limb darkening parameters
for a  [Fe/H]$=-1$ dex or  [Fe/H]$=-3$ dex star changes  the resulting
mass and radius values by 10 and  15 per cent from the [Fe/H]$=-2$ dex
parameters (note  that only  these three  [Fe/H] abundance  values are
available    for    acquiring    the    limb-darkening    coefficients
\cite{claret2011}), meaning that fits relying  on the amplitude of the
ellipsoidal modulation are quite uncertain.

We managed  to break the degeneracy  between the scaled radii  and the
inclination  by measuring  the  depth of  the  secondary eclipse  (see
Figure~\ref{fig:full_lc}),  an approach  that  allowed  us to  measure
precise parameters for the stars. It has to be stressed that following
this procedure considerably alleviates  the dependence of the measured
parameters with  the adopted limb-darkening coefficients  for the cool
subdwarf star, i.e.  the resulting masses and radii change by only 1.5
and 2 per cent in this case  depending on the adopted value of [Fe/H],
comparable  to our  measured  uncertainties. However,  the  lack of  a
direct  determination of  the  radial velocity  semi-amplitude of  the
white dwarf $K_\mathrm{WD}$  implied that we were not  able to measure
the  masses  and radii  of  the  white  dwarf  and the  cool  subdwarf
independently  of  any  theoretical   model.   Fortunately,  a  recent
observational  study shows  that  the white  dwarf  measured radii  in
eclipsing  binaries  are  in   excellent  agreement  with  theoretical
expectations  \cite{parsons2017}.   Therefore,  we adopted  the  white
dwarf's radius to follow  the theoretical mass-radius relationship for
a  13,250\,K helium-core  white dwarf  with a  thick surface  hydrogen
envelope \cite{panei2007} when fitting  the HiPERCAM light curves.  We
allowed  some departure  from the  relation  to account  for a  500\,K
uncertainty in  the white dwarf  temperature.  Note that the  radii of
very low  metallicity ([Fe/H]<-3  dex) white dwarfs  are 1-2  per cent
larger than solar metallicity white dwarfs\cite{Althaus2017}, which is
comparable to  the departure  allowed to  account for  the temperature
uncertainty.

For determining the distributions of  our model parameters we employed
the Markov Chain Monte  Carlo (MCMC) method\cite{press2007}.  We based
the likelihood of  accepting a model on a combination  of the $\chi^2$
of the light-curve fit to the data and an additional prior probability
whereby  we  ensure  that   $k_\mathrm{sd}$  is  consistent  with  the
spectroscopic result and  that the white dwarf's  radius is consistent
with the mass-radius relationship outlined  above.  For each band, the
approximate best  parameters and covariances were  determined using an
initial  MCMC  chain.  These  approximate  best  parameters were  then
considered  as  the  starting  values for  longer  chains  which  were
employed to derive the final  model values and their uncertainties. We
simultaneously run four chains in  order to ensure that they converged
on  the same  values, with  the initial  approximated best  parameters
slightly  perturbed. From  each chain,  the 50,000  first points  were
classified as a ``burn-in'' phase  and were removed for the subsequent
analysis. Each chain had a total length of 250,000 points.

The   full    light   curves   and    model   fits   are    shown   in
Figure~\ref{fig:full_lc}. The  parameters that  result from  the $u_s$
and $g_s$  band fits have  associated large uncertainties  since these
bands are  dominated by  the white dwarf'  flux and  therefore contain
relatively little information regarding the inclination. The secondary
eclipse is only detected in the  $i_s$ band and therefore the tightest
constraints come from these data.

It is  important to  emphasise the  light-curve fitting  code provides
both the  tidally-distorted radius of  the cool subdwarf  (towards the
white    dwarf)    as    well   as    its    volume-averaged    radius
(Table\,\ref{tab:params}).  We adopted  the  latter  in our  analysis,
which is expected  to make the effects of  tidal distortion negligible
when comparing the observed to the theoretical relations for this star
\cite{Parsons2018Md}.

\subsection*{The VOSA spectral energy distribution fitting routine}
\label{m:vosa}

Different statistical tests  may be employed by  VOSA for determining
which model  fits best the observational  data.  In our case,  we used
the  $\chi^2$  statistical test  between  the  model spectra  and  the
observed photometry to derive the effective temperature and luminosity
of  the cool  subdwarf.   Good  fits are  achieved  by  VOSA when  the
parameter Vgfb\,$<$\,10--15.  Vfgb is  defined as the modified reduced
$\chi^{2}$ and it is  calculated by forcing $\sigma(F_{obs})>0.1\times
F_{obs}$,  where  $\sigma(F_{obs})$ is  the  error  of $F_{obs}$  (the
observed flux).

The SED  was compared to  three different grids of  synthetic spectra,
namely   BT-Settl,  BT-NextGen   (AGSS2009)  and   BT-NextGen  (GNS93)
\cite{Allard2012}.   We  note that  we  did  not use  other  available
libraries within VOSA due to the  lack of models at low metallicites.
During   the  fitting   we  accounted   for  interstellar   extinction
\cite{Schlafly+Finkbeiner11}   and   used    the   Bayesian   distance
\cite{Bailer2018} derived  from the  measured $Gaia$ parallax  and its
uncertainty  \cite{Gaia2018}  as   the  distance  (541$^{+216}_{-124}$
pc). We fixed  the surface gravity to be within  the range 5--5.5 dex,
thus covering the 5.16 dex value that results from our mass and radius
determinations  from the  light curve  fit.  Finally,  the metallicity
(i.e. Fe/H abundance) was assumed to  be lower than -1 dex, consistent
with that of  an usd star.  In all cases  Vgfb$<15$, with our best-fit
model  achieving Vgfb$=6.6$,  which corroborates  the goodness  of our
fit.

VOSA  uses the  bolometric  flux  and the  distance  to determine  the
luminosity. The bolometric flux is obtained integrating the flux using
the observational  photometric points. In  cases where the SED  is not
covered  by  the  observational  photometry, VOSA  uses  the  best-fit
theoretical model.

\subsection*{The iSpec [Fe/H] determination}
\label{m:ispec}

The  metallicity, in  particular  the [Fe/H]  abundance,  of the  cool
subdwarf of  SDSS\,J2355+0448 was  additionally estimated  using iSpec
\cite{Blanco2014}, a  code for the  treatment and analysis  of stellar
spectra. iSpec  allows determining the atmospheric  stellar parameters
(including  the effective  temperature, surface  gravity, metallicity,
micro-  and   macro-turbulence,  rotation)  and   individual  chemical
abundances for FGKM stars via two different methods: (1) the synthetic
spectral fitting technique and (2) the equivalent widths method. Here,
we  adopted  the  synthetic   spectral  fitting  technique.   This  is
implemented  by iSpec  via  minimizing the  $\chi^2$  between a  given
observed spectrum  and an  adopted library  of synthetic  spectra.  We
applied  the  synthetic spectral  fitting  technique  to the  observed
(combined   and  white   dwarf  subtracted)   X-Shooter  spectrum   of
SDSS\,J2355+0448.

iSpec requires the  user to mask the lines  and corresponding segments
to  be used  in the  fitting.  To that  end we  used the  \Ion{Fe}{I},
\Ion{Fe}{II},  \Ion{Ca}{I},  \Ion{Ca}{II}, \Ion{Na}{I},  \Ion{Na}{II},
\Ion{K}{I},   \Ion{K}{II}   absorption   lines  located   within   the
5500--9000\AA\,  wavelength range  sampled by  the combined  and white
dwarf subtracted X-Shooter VIS arm  spectrum. To do the computation of
the  $\chi^2$, iSpec  used the  GES  ($Gaia$ ESO  survey) atomic  line
lists.   The  synthetic   spectra  were   computed  using   the  MARCS
\cite{Gustafsson2008} model  atmosphere library and the  SPECTRUM code
within  iSpec.  An  initial  set of  atmospheric  parameters  is  also
required prior to the fitting. We used \teff\,=\,3650\,K (based on our
VOSA  fitting results),  \logg\,=\,5.16\,dex  (based on  the mass  and
radius determined from the light  curve fitting), [M/H] $=-1.55$ \,dex
(we    used     the    [Fe/H]     abundance    derived     from    the
$\tau_\mathrm{TiO/CaH}=0.095$  ratio  as  a  proxy  for  metallicity),
[$\alpha/$Fe]\,=\,0\,dex,           micro-turbulence          velocity
$V_\mathrm{mic}$\,=\,1.05\,\kms,       macro-turbulence       velocity
$V_\mathrm{mac}$\,=\,4.21\,\kms,          $v$\,sin($i$)\,=\,94.3\,\kms
(obtained from the measured orbital period of the binary and radius of
the cool subdwarf,  together with the known  binary inclination), limb
darkening  coefficient\,=\,0.6,  resolution  R\,=\,8,900 (set  by  our
X-Shooter  VIS spectrum)  and radial  velocity\,=\,0\,\kms (since  the
rest-frame  radial  velocity  correction   was  performed  before  the
fitting). The  initial guesses of  \teff, \logg, $v$\,sin($i$),  R and
radial velocity were fixed during  the fitting process and the maximum
number  of iterations  was set  to  10. This  resulted in  a value  of
[Fe/H]$=-2.1 \pm 2$ dex for the cool subdwarf star.

\subsection*{The binary$\_$c stellar evolution code}
\label{m:binaryc}

We predict  the future evolution  of SDSS\,J2355+0448 by means  of the
binary        stellar         evolution        code        binary$\_$c
\cite{Izzard2004,Izzard2018}, a  software for the evolution  of single
and   binary   stars,   calculations  of   stellar   populations   and
nucleosynthesis analysis.   binary$\_$c updates  and enhances  the BSE
(Binary  Star  Evolution)  code \cite{Hurley2002},  providing  stellar
evolution for  stars in the  range 0.1 to  100\,M$_\mathrm{\odot}$. In
order to  simulate the evolution  of a given binary  star, binary$\_$c
requires a  set of initial parameters  such as the mass,  stellar type
(i.e.   main  sequence  star,  white  dwarf,  etc.),  metallicity  and
rotational  velocity of  each star,  as  well as  the orbital  period,
eccentricity  and orbital  separation of  the binary.   In all  cases,
except  for  the  rotational  velocity  of the  white  dwarf  and  the
metallicity of the stars, the input parameters are tightly constrained
from our  observations (note  that the eccentricity  must be  zero for
such a  close binary star)  and we hence  used the observed  values as
input parameters in  the simulation. We assumed  a rotational velocity
of 0 \kms\, for the white  dwarf and we considered an [Fe/H] abundance
of -1.55 dex for the cool subdwarf.  Additional parameters/assumptions
are required by binary$\_$c regarding  explosions (not relevant in our
case),  Roche-lobe overflow,  common  envelope  efficiency, winds  and
nucleosynthesis (also  not relevant in  our case).  In those  cases we
assumed  the standard  values  as provided  by  binary$\_$c.  We  note
however that modifying these assumptions had no dramatic effect in the
evolutionary  paths  expected   for  SDSS\,J2355+0448,  except  slight
changes in the evolutionary timescales.

\section*{Code availability statement}

The spectral/decomposition routine and the radial-velocity fitting
method used in this work are not publicly available. The light-curve
fitting method is available at
\url{https://github.com/trmrsh/cpp-lcurve}.  MIDAS is available at
\url{https://www.eso.org/sci/software/esomidas//}.  VOSA is available
at \url{http://svo2.cab.inta-csic.es/theory/vosa/}. iSpec is available
at \url{https://www.blancocuaresma.com/s/iSpec}. The binary\_c stellar
evolution code is available at
\url{https://www.ast.cam.ac.uk/~rgi/binary\_c.html}. The X-Shooter
reduction pipeline (version 2.9.3) is available at
\url{https://www.eso.org/sci/software/pipelines/xshooter/} and the
dedicated HiPERCAM pipeline at
\url{https://github.com/HiPERCAM/hipercam}.

\section*{Data availability statement}

Based on  observations made with  the Gran Telescopio  Canarias (GTC),
installed in  the Spanish Observatorio  del Roque de los  Muchachos of
the Instituto de Astrof\'isica de Canarias,  in the island of La Palma
(program GTC21-18B). Based on observations made with ESO Telescopes at
the La Silla Paranal Observatory under programme ID\,2100.D-5022(A).

This  publication  makes use  of  VOSA,  developed under  the  Spanish
Virtual Observatory project supported  from the Spanish MINECO through
grant AyA2017-84089. This work has made  use of data from the European
Space        Agency       (ESA)        mission       {\it        Gaia}
(\url{https://www.cosmos.esa.int/gaia}), processed  by the  {\it Gaia}
Data      Processing      and     Analysis      Consortium      (DPAC,
\url{https://www.cosmos.esa.int/web/gaia/dpac/consortium}).

Figures 1, 3 and 4 and Supplementary Figures 1, 2 and 4 have associated
raw data.  The data that support the plots within this paper and other
findings of  this study  are available  from the  corresponding author
upon reasonable request.


\begin{thebibliography}{10}
\urlstyle{rm}
\expandafter\ifx\csname url\endcsname\relax
  \def\url#1{\texttt{#1}}\fi
\expandafter\ifx\csname urlprefix\endcsname\relax\def\urlprefix{URL }\fi
\expandafter\ifx\csname doiprefix\endcsname\relax\def\doiprefix{DOI: }\fi
\providecommand{\bibinfo}[2]{#2}
\providecommand{\eprint}[2][]{\url{#2}}

\bibitem{Burrows1993}
\bibinfo{author}{{Burrows}, A.}, \bibinfo{author}{{Hubbard}, W.~B.},
  \bibinfo{author}{{Saumon}, D.} \& \bibinfo{author}{{Lunine}, J.~I.}
\newblock \bibinfo{journal}{\bibinfo{title}{{An expanded set of brown dwarf and
  very low mass star models}}}.
\newblock {\emph{\JournalTitle{\apj}}} \textbf{\bibinfo{volume}{406}},
  \bibinfo{pages}{158--171}, \doiprefix\url{10.1086/172427}
  (\bibinfo{year}{1993}).

\bibitem{Morgan1943}
\bibinfo{author}{{Morgan}, W.~W.}, \bibinfo{author}{{Keenan}, P.~C.} \&
  \bibinfo{author}{{Kellman}, E.}
\newblock \emph{\bibinfo{title}{{An atlas of stellar spectra, with an outline
  of spectral classification}}} (\bibinfo{publisher}{The University of Chicago
  press}, \bibinfo{year}{1943}).

\bibitem{Burgasser2003}
\bibinfo{author}{{Burgasser}, A.~J.} \emph{et~al.}
\newblock \bibinfo{journal}{\bibinfo{title}{{The First Substellar Subdwarf?
  Discovery of a Metal-poor L Dwarf with Halo Kinematics}}}.
\newblock {\emph{\JournalTitle{\apj}}} \textbf{\bibinfo{volume}{592}},
  \bibinfo{pages}{1186--1192}, \doiprefix\url{10.1086/375813}
  (\bibinfo{year}{2003}).
\newblock \eprint{astro-ph/0304174}.

\bibitem{Digby2003}
\bibinfo{author}{{Digby}, A.~P.}, \bibinfo{author}{{Hambly}, N.~C.},
  \bibinfo{author}{{Cooke}, J.~A.}, \bibinfo{author}{{Reid}, I.~N.} \&
  \bibinfo{author}{{Cannon}, R.~D.}
\newblock \bibinfo{journal}{\bibinfo{title}{{The subdwarf luminosity
  function}}}.
\newblock {\emph{\JournalTitle{\mnras}}} \textbf{\bibinfo{volume}{344}},
  \bibinfo{pages}{583--601}, \doiprefix\url{10.1046/j.1365-8711.2003.06842.x}
  (\bibinfo{year}{2003}).
\newblock \eprint{astro-ph/0304056}.

\bibitem{Rajpurohit2014}
\bibinfo{author}{{Rajpurohit}, A.~S.} \emph{et~al.}
\newblock \bibinfo{journal}{\bibinfo{title}{{High-resolution spectroscopic
  atlas of M subdwarfs. Effective temperature and metallicity}}}.
\newblock {\emph{\JournalTitle{\aap}}} \textbf{\bibinfo{volume}{564}},
  \bibinfo{pages}{A90}, \doiprefix\url{10.1051/0004-6361/201322881}
  (\bibinfo{year}{2014}).
\newblock \eprint{1401.2901}.

\bibitem{Lepine2007}
\bibinfo{author}{{L{\'e}pine}, S.}, \bibinfo{author}{{Rich}, R.~M.} \&
  \bibinfo{author}{{Shara}, M.~M.}
\newblock \bibinfo{journal}{\bibinfo{title}{{Revised Metallicity Classes for
  Low-Mass Stars: Dwarfs (dM), Subdwarfs (sdM), Extreme Subdwarfs (esdM), and
  Ultrasubdwarfs (usdM)}}}.
\newblock {\emph{\JournalTitle{\apj}}} \textbf{\bibinfo{volume}{669}},
  \bibinfo{pages}{1235--1247}, \doiprefix\url{10.1086/521614}
  (\bibinfo{year}{2007}).
\newblock \eprint{0707.2993}.

\bibitem{Woolf2009}
\bibinfo{author}{{Woolf}, V.~M.}, \bibinfo{author}{{L{\'e}pine}, S.} \&
  \bibinfo{author}{{Wallerstein}, G.}
\newblock \bibinfo{journal}{\bibinfo{title}{{Calibrating M-Dwarf Metallicities
  Using Molecular Indices: Extension to Low-metallicity Stars}}}.
\newblock {\emph{\JournalTitle{\pasp}}} \textbf{\bibinfo{volume}{121}},
  \bibinfo{pages}{117}, \doiprefix\url{10.1086/597433} (\bibinfo{year}{2009}).

\bibitem{Gizis1997}
\bibinfo{author}{{Gizis}, J.~E.}
\newblock \bibinfo{journal}{\bibinfo{title}{{M-Subdwarfs: Spectroscopic
  Classification and the Metallicity Scale}}}.
\newblock {\emph{\JournalTitle{\aj}}} \textbf{\bibinfo{volume}{113}},
  \bibinfo{pages}{806--822}, \doiprefix\url{10.1086/118302}
  (\bibinfo{year}{1997}).
\newblock \eprint{astro-ph/9611222}.

\bibitem{Stoughton2002}
\bibinfo{author}{{Stoughton}, C.} \emph{et~al.}
\newblock \bibinfo{journal}{\bibinfo{title}{{Sloan Digital Sky Survey: Early
  Data Release}}}.
\newblock {\emph{\JournalTitle{\aj}}} \textbf{\bibinfo{volume}{123}},
  \bibinfo{pages}{485--548}, \doiprefix\url{10.1086/324741}
  (\bibinfo{year}{2002}).

\bibitem{Cui2012}
\bibinfo{author}{{Cui}, X.-Q.} \emph{et~al.}
\newblock \bibinfo{journal}{\bibinfo{title}{{The Large Sky Area Multi-Object
  Fiber Spectroscopic Telescope (LAMOST)}}}.
\newblock {\emph{\JournalTitle{Research in Astronomy and Astrophysics}}}
  \textbf{\bibinfo{volume}{12}}, \bibinfo{pages}{1197--1242},
  \doiprefix\url{10.1088/1674-4527/12/9/003} (\bibinfo{year}{2012}).

\bibitem{Savcheva2014}
\bibinfo{author}{{Savcheva}, A.~S.}, \bibinfo{author}{{West}, A.~A.} \&
  \bibinfo{author}{{Bochanski}, J.~J.}
\newblock \bibinfo{journal}{\bibinfo{title}{{A New Sample of Cool Subdwarfs
  from SDSS: Properties and Kinematics}}}.
\newblock {\emph{\JournalTitle{\apj}}} \textbf{\bibinfo{volume}{794}},
  \bibinfo{pages}{145}, \doiprefix\url{10.1088/0004-637X/794/2/145}
  (\bibinfo{year}{2014}).
\newblock \eprint{1409.1229}.

\bibitem{Bai2016}
\bibinfo{author}{{Bai}, Y.} \emph{et~al.}
\newblock \bibinfo{journal}{\bibinfo{title}{{Nearby M subdwarfs from LAMOST
  data release 2}}}.
\newblock {\emph{\JournalTitle{Research in Astronomy and Astrophysics}}}
  \textbf{\bibinfo{volume}{16}}, \bibinfo{pages}{107},
  \doiprefix\url{10.1088/1674-4527/16/7/107} (\bibinfo{year}{2016}).

\bibitem{Kesseli2018}
\bibinfo{author}{{Kesseli}, A.~Y.} \emph{et~al.}
\newblock \bibinfo{journal}{\bibinfo{title}{{Radii of 88 M Subdwarfs and
  Updated Radius Relations for Low-Metallicity M Dwarf Stars}}}.
\newblock {\emph{\JournalTitle{ArXiv e-prints}}}  (\bibinfo{year}{2018}).
\newblock \eprint{1810.07702}.

\bibitem{Jao2016}
\bibinfo{author}{{Jao}, W.-C.}, \bibinfo{author}{{Nelan}, E.~P.},
  \bibinfo{author}{{Henry}, T.~J.}, \bibinfo{author}{{Franz}, O.~G.} \&
  \bibinfo{author}{{Wasserman}, L.~H.}
\newblock \bibinfo{journal}{\bibinfo{title}{{Cool Subdwarf Investigations. III.
  Dynamical Masses of Low-metallicity Subdwarfs}}}.
\newblock {\emph{\JournalTitle{\aj}}} \textbf{\bibinfo{volume}{152}},
  \bibinfo{pages}{153}, \doiprefix\url{10.3847/0004-6256/152/6/153}
  (\bibinfo{year}{2016}).
\newblock \eprint{1607.01304}.

\bibitem{Kepler2015}
\bibinfo{author}{{Kepler}, S.~O.} \emph{et~al.}
\newblock \bibinfo{journal}{\bibinfo{title}{{New white dwarf stars in the Sloan
  Digital Sky Survey Data Release 10}}}.
\newblock {\emph{\JournalTitle{\mnras}}} \textbf{\bibinfo{volume}{446}},
  \bibinfo{pages}{4078--4087}, \doiprefix\url{10.1093/mnras/stu2388}
  (\bibinfo{year}{2015}).
\newblock \eprint{1411.4149}.

\bibitem{Rebassa2016}
\bibinfo{author}{{Rebassa-Mansergas}, A.} \emph{et~al.}
\newblock \bibinfo{journal}{\bibinfo{title}{{The SDSS spectroscopic catalogue
  of white dwarf-main-sequence binaries: new identifications from DR 9-12}}}.
\newblock {\emph{\JournalTitle{\mnras}}} \textbf{\bibinfo{volume}{458}},
  \bibinfo{pages}{3808--3819}, \doiprefix\url{10.1093/mnras/stw554}
  (\bibinfo{year}{2016}).
\newblock \eprint{1603.01017}.

\bibitem{Ren2018}
\bibinfo{author}{{Ren}, J.-J.} \emph{et~al.}
\newblock \bibinfo{journal}{\bibinfo{title}{{White dwarf-main sequence binaries
  from LAMOST: the DR5 catalogue}}}.
\newblock {\emph{\JournalTitle{\mnras}}} \textbf{\bibinfo{volume}{477}},
  \bibinfo{pages}{4641--4654}, \doiprefix\url{10.1093/mnras/sty805}
  (\bibinfo{year}{2018}).
\newblock \eprint{1803.09523}.

\bibitem{parsons2017}
\bibinfo{author}{{Parsons}, S.~G.} \emph{et~al.}
\newblock \bibinfo{journal}{\bibinfo{title}{{Testing the white dwarf
  mass-radius relationship with eclipsing binaries}}}.
\newblock {\emph{\JournalTitle{\mnras}}} \textbf{\bibinfo{volume}{470}},
  \bibinfo{pages}{4473--4492}, \doiprefix\url{10.1093/mnras/stx1522}
  (\bibinfo{year}{2017}).
\newblock \eprint{1706.05016}.

\bibitem{parsons2018}
\bibinfo{author}{{Parsons}, S.~G.} \emph{et~al.}
\newblock \bibinfo{journal}{\bibinfo{title}{{The scatter of the M dwarf
  mass-radius relationship}}}.
\newblock {\emph{\JournalTitle{\mnras}}} \textbf{\bibinfo{volume}{481}},
  \bibinfo{pages}{1083--1096}, \doiprefix\url{10.1093/mnras/sty2345}
  (\bibinfo{year}{2018}).
\newblock \eprint{1808.07780}.

\bibitem{Drake2009}
\bibinfo{author}{{Drake}, A.~J.} \emph{et~al.}
\newblock \bibinfo{journal}{\bibinfo{title}{{First Results from the Catalina
  Real-Time Transient Survey}}}.
\newblock {\emph{\JournalTitle{\apj}}} \textbf{\bibinfo{volume}{696}},
  \bibinfo{pages}{870--884}, \doiprefix\url{10.1088/0004-637X/696/1/870}
  (\bibinfo{year}{2009}).
\newblock \eprint{0809.1394}.

\bibitem{dhillon2018}
\bibinfo{author}{{Dhillon}, V.} \emph{et~al.}
\newblock \bibinfo{title}{{First light with HiPERCAM on the GTC}}.
\newblock In \emph{\bibinfo{booktitle}{Society of Photo-Optical Instrumentation
  Engineers (SPIE) Conference Series}}, vol. \bibinfo{volume}{10702} of
  \emph{\bibinfo{series}{Society of Photo-Optical Instrumentation Engineers
  (SPIE) Conference Series}}, \bibinfo{pages}{107020L},
  \doiprefix\url{10.1117/12.2312041} (\bibinfo{year}{2018}).
\newblock \eprint{1807.00557}.

\bibitem{Rebassa2007}
\bibinfo{author}{{Rebassa-Mansergas}, A.}, \bibinfo{author}{{G{\"a}nsicke},
  B.~T.}, \bibinfo{author}{{Rodr{\'{\i}}guez-Gil}, P.},
  \bibinfo{author}{{Schreiber}, M.~R.} \& \bibinfo{author}{{Koester}, D.}
\newblock \bibinfo{journal}{\bibinfo{title}{{Post-common-envelope binaries from
  SDSS - I. 101 white dwarf main-sequence binaries with multiple Sloan Digital
  Sky Survey spectroscopy}}}.
\newblock {\emph{\JournalTitle{\mnras}}} \textbf{\bibinfo{volume}{382}},
  \bibinfo{pages}{1377--1393}, \doiprefix\url{10.1111/j.1365-2966.2007.12288.x}
  (\bibinfo{year}{2007}).
\newblock \eprint{0707.4107}.

\bibitem{Rajpurohit2016}
\bibinfo{author}{{Rajpurohit}, A.~S.} \emph{et~al.}
\newblock \bibinfo{journal}{\bibinfo{title}{{Spectral energy distribution of
  M-subdwarfs: A study of their atmospheric properties}}}.
\newblock {\emph{\JournalTitle{\aap}}} \textbf{\bibinfo{volume}{596}},
  \bibinfo{pages}{A33}, \doiprefix\url{10.1051/0004-6361/201526776}
  (\bibinfo{year}{2016}).
\newblock \eprint{1609.07062}.

\bibitem{Koester2010}
\bibinfo{author}{{Koester}, D.}
\newblock \bibinfo{journal}{\bibinfo{title}{White dwarf spectra and atmosphere
  models}}.
\newblock {\emph{\JournalTitle{Memorie della Societa Astronomica Italiana}}}
  \textbf{\bibinfo{volume}{81}}, \bibinfo{pages}{921--931}
  (\bibinfo{year}{2010}).

\bibitem{Bailer2018}
\bibinfo{author}{{Bailer-Jones}, C.~A.~L.}, \bibinfo{author}{{Rybizki}, J.},
  \bibinfo{author}{{Fouesneau}, M.}, \bibinfo{author}{{Mantelet}, G.} \&
  \bibinfo{author}{{Andrae}, R.}
\newblock \bibinfo{journal}{\bibinfo{title}{{Estimating Distance from
  Parallaxes. IV. Distances to 1.33 Billion Stars in Gaia Data Release 2}}}.
\newblock {\emph{\JournalTitle{\aj}}} \textbf{\bibinfo{volume}{156}},
  \bibinfo{pages}{58}, \doiprefix\url{10.3847/1538-3881/aacb21}
  (\bibinfo{year}{2018}).
\newblock \eprint{1804.10121}.

\bibitem{Althaus2017}
\bibinfo{author}{{Althaus}, L.~G.}, \bibinfo{author}{{De Ger{\'o}nimo}, F.},
  \bibinfo{author}{{C{\'o}rsico}, A.}, \bibinfo{author}{{Torres}, S.} \&
  \bibinfo{author}{{Garc{\'{\i}}a-Berro}, E.}
\newblock \bibinfo{journal}{\bibinfo{title}{{The evolution of white dwarfs
  resulting from helium-enhanced, low-metallicity progenitor stars}}}.
\newblock {\emph{\JournalTitle{\aap}}} \textbf{\bibinfo{volume}{597}},
  \bibinfo{pages}{A67}, \doiprefix\url{10.1051/0004-6361/201629909}
  (\bibinfo{year}{2017}).
\newblock \eprint{1611.06191}.

\bibitem{copperwheat2010}
\bibinfo{author}{{Copperwheat}, C.~M.} \emph{et~al.}
\newblock \bibinfo{journal}{\bibinfo{title}{{Physical properties of IP Pegasi:
  an eclipsing dwarf nova with an unusually cool white dwarf}}}.
\newblock {\emph{\JournalTitle{\mnras}}} \textbf{\bibinfo{volume}{402}},
  \bibinfo{pages}{1824--1840}, \doiprefix\url{10.1111/j.1365-2966.2009.16010.x}
  (\bibinfo{year}{2010}).
\newblock \eprint{0911.1637}.

\bibitem{Bayo08}
\bibinfo{author}{{Bayo}, A.} \emph{et~al.}
\newblock \bibinfo{journal}{\bibinfo{title}{{VOSA: virtual observatory SED
  analyzer. An application to the Collinder 69 open cluster}}}.
\newblock {\emph{\JournalTitle{\aap}}} \textbf{\bibinfo{volume}{492}},
  \bibinfo{pages}{277--287}, \doiprefix\url{10.1051/0004-6361:200810395}
  (\bibinfo{year}{2008}).
\newblock \eprint{0808.0270}.

\bibitem{Pickles1998}
\bibinfo{author}{{Pickles}, A.~J.}
\newblock \bibinfo{journal}{\bibinfo{title}{{A Stellar Spectral Flux Library:
  1150-25000 {\AA}}}}.
\newblock {\emph{\JournalTitle{\pasp}}} \textbf{\bibinfo{volume}{110}},
  \bibinfo{pages}{863--878}, \doiprefix\url{10.1086/316197}
  (\bibinfo{year}{1998}).

\bibitem{Hewett06}
\bibinfo{author}{{Hewett}, P.~C.}, \bibinfo{author}{{Warren}, S.~J.},
  \bibinfo{author}{{Leggett}, S.~K.} \& \bibinfo{author}{{Hodgkin}, S.~T.}
\newblock \bibinfo{journal}{\bibinfo{title}{{The UKIRT Infrared Deep Sky Survey
  ZY JHK photometric system: passbands and synthetic colours}}}.
\newblock {\emph{\JournalTitle{\mnras}}} \textbf{\bibinfo{volume}{367}},
  \bibinfo{pages}{454--468}, \doiprefix\url{10.1111/j.1365-2966.2005.09969.x}
  (\bibinfo{year}{2006}).
\newblock \eprint{astro-ph/0601592}.

\bibitem{Wright10}
\bibinfo{author}{{Wright}, E.~L.} \emph{et~al.}
\newblock \bibinfo{journal}{\bibinfo{title}{{The Wide-field Infrared Survey
  Explorer (WISE): Mission Description and Initial On-orbit Performance}}}.
\newblock {\emph{\JournalTitle{\aj}}} \textbf{\bibinfo{volume}{140}},
  \bibinfo{pages}{1868--1881}, \doiprefix\url{10.1088/0004-6256/140/6/1868}
  (\bibinfo{year}{2010}).
\newblock \eprint{1008.0031}.

\bibitem{Skrutskie06}
\bibinfo{author}{{Skrutskie}, M.~F.} \emph{et~al.}
\newblock \bibinfo{journal}{\bibinfo{title}{{The Two Micron All Sky Survey
  (2MASS)}}}.
\newblock {\emph{\JournalTitle{\aj}}} \textbf{\bibinfo{volume}{131}},
  \bibinfo{pages}{1163--1183}, \doiprefix\url{10.1086/498708}
  (\bibinfo{year}{2006}).

\bibitem{Cross12}
\bibinfo{author}{{Cross}, N.~J.~G.} \emph{et~al.}
\newblock \bibinfo{journal}{\bibinfo{title}{{The VISTA Science Archive}}}.
\newblock {\emph{\JournalTitle{\aap}}} \textbf{\bibinfo{volume}{548}},
  \bibinfo{pages}{A119}, \doiprefix\url{10.1051/0004-6361/201219505}
  (\bibinfo{year}{2012}).
\newblock \eprint{1210.2980}.

\bibitem{Lepine2013}
\bibinfo{author}{{L{\'e}pine}, S.} \emph{et~al.}
\newblock \bibinfo{journal}{\bibinfo{title}{{A Spectroscopic Catalog of the
  Brightest (J $<$ 9) M Dwarfs in the Northern Sky}}}.
\newblock {\emph{\JournalTitle{\aj}}} \textbf{\bibinfo{volume}{145}},
  \bibinfo{pages}{102}, \doiprefix\url{10.1088/0004-6256/145/4/102}
  (\bibinfo{year}{2013}).
\newblock \eprint{1206.5991}.

\bibitem{Mann2013}
\bibinfo{author}{{Mann}, A.~W.}, \bibinfo{author}{{Brewer}, J.~M.},
  \bibinfo{author}{{Gaidos}, E.}, \bibinfo{author}{{L{\'e}pine}, S.} \&
  \bibinfo{author}{{Hilton}, E.~J.}
\newblock \bibinfo{journal}{\bibinfo{title}{{Prospecting in Late-type Dwarfs: A
  Calibration of Infrared and Visible Spectroscopic Metallicities of Late K and
  M Dwarfs Spanning 1.5 dex}}}.
\newblock {\emph{\JournalTitle{\aj}}} \textbf{\bibinfo{volume}{145}},
  \bibinfo{pages}{52}, \doiprefix\url{10.1088/0004-6256/145/2/52}
  (\bibinfo{year}{2013}).
\newblock \eprint{1211.4630}.

\bibitem{Blanco2014}
\bibinfo{author}{{Blanco-Cuaresma}, S.}, \bibinfo{author}{{Soubiran}, C.},
  \bibinfo{author}{{Heiter}, U.} \& \bibinfo{author}{{Jofr{\'e}}, P.}
\newblock \bibinfo{journal}{\bibinfo{title}{{Determining stellar atmospheric
  parameters and chemical abundances of FGK stars with iSpec}}}.
\newblock {\emph{\JournalTitle{\aap}}} \textbf{\bibinfo{volume}{569}},
  \bibinfo{pages}{A111}, \doiprefix\url{10.1051/0004-6361/201423945}
  (\bibinfo{year}{2014}).
\newblock \eprint{1407.2608}.

\bibitem{iben+livio1993}
\bibinfo{author}{{Iben}, I., Jr.} \& \bibinfo{author}{{Livio}, M.}
\newblock \bibinfo{journal}{\bibinfo{title}{{Common envelopes in binary star
  evolution}}}.
\newblock {\emph{\JournalTitle{\pasp}}} \textbf{\bibinfo{volume}{105}},
  \bibinfo{pages}{1373--1406}, \doiprefix\url{10.1086/133321}
  (\bibinfo{year}{1993}).

\bibitem{Hurley2002}
\bibinfo{author}{{Hurley}, J.~R.}, \bibinfo{author}{{Tout}, C.~A.} \&
  \bibinfo{author}{{Pols}, O.~R.}
\newblock \bibinfo{journal}{\bibinfo{title}{{Evolution of binary stars and the
  effect of tides on binary populations}}}.
\newblock {\emph{\JournalTitle{\mnras}}} \textbf{\bibinfo{volume}{329}},
  \bibinfo{pages}{897--928}, \doiprefix\url{10.1046/j.1365-8711.2002.05038.x}
  (\bibinfo{year}{2002}).
\newblock \eprint{astro-ph/0201220}.

\bibitem{Parsons2018Md}
\bibinfo{author}{{Parsons}, S.~G.} \emph{et~al.}
\newblock \bibinfo{journal}{\bibinfo{title}{{The scatter of the M dwarf
  mass-radius relationship}}}.
\newblock {\emph{\JournalTitle{\mnras}}} \textbf{\bibinfo{volume}{481}},
  \bibinfo{pages}{1083--1096}, \doiprefix\url{10.1093/mnras/sty2345}
  (\bibinfo{year}{2018}).
\newblock \eprint{1808.07780}.

\bibitem{Catalan2008}
\bibinfo{author}{{Catal{\'a}n}, S.}, \bibinfo{author}{{Isern}, J.},
  \bibinfo{author}{{Garc{\'{\i}}a-Berro}, E.} \& \bibinfo{author}{{Ribas}, I.}
\newblock \bibinfo{journal}{\bibinfo{title}{{The initial-final mass
  relationship of white dwarfs revisited: effect on the luminosity function and
  mass distribution}}}.
\newblock {\emph{\JournalTitle{\mnras}}} \textbf{\bibinfo{volume}{387}},
  \bibinfo{pages}{1693--1706}, \doiprefix\url{10.1111/j.1365-2966.2008.13356.x}
  (\bibinfo{year}{2008}).
\newblock \eprint{0804.3034}.

\bibitem{Cummings2018}
\bibinfo{author}{{Cummings}, J.~D.}, \bibinfo{author}{{Kalirai}, J.~S.},
  \bibinfo{author}{{Tremblay}, P.-E.}, \bibinfo{author}{{Ramirez-Ruiz}, E.} \&
  \bibinfo{author}{{Choi}, J.}
\newblock \bibinfo{journal}{\bibinfo{title}{{The White Dwarf Initial-Final Mass
  Relation for Progenitor Stars from 0.85 to 7.5 M $_{\odot}$}}}.
\newblock {\emph{\JournalTitle{\apj}}} \textbf{\bibinfo{volume}{866}},
  \bibinfo{pages}{21}, \doiprefix\url{10.3847/1538-4357/aadfd6}
  (\bibinfo{year}{2018}).
\newblock \eprint{1809.01673}.

\bibitem{Izzard2004}
\bibinfo{author}{{Izzard}, R.~G.}, \bibinfo{author}{{Tout}, C.~A.},
  \bibinfo{author}{{Karakas}, A.~I.} \& \bibinfo{author}{{Pols}, O.~R.}
\newblock \bibinfo{journal}{\bibinfo{title}{{A new synthetic model for
  asymptotic giant branch stars}}}.
\newblock {\emph{\JournalTitle{\mnras}}} \textbf{\bibinfo{volume}{350}},
  \bibinfo{pages}{407--426}, \doiprefix\url{10.1111/j.1365-2966.2004.07446.x}
  (\bibinfo{year}{2004}).
\newblock \eprint{astro-ph/0402403}.

\bibitem{Izzard2018}
\bibinfo{author}{{Izzard}, R.~G.} \emph{et~al.}
\newblock \bibinfo{journal}{\bibinfo{title}{{Binary stars in the Galactic thick
  disc}}}.
\newblock {\emph{\JournalTitle{\mnras}}} \textbf{\bibinfo{volume}{473}},
  \bibinfo{pages}{2984--2999}, \doiprefix\url{10.1093/mnras/stx2355}
  (\bibinfo{year}{2018}).
\newblock \eprint{1709.05237}.

\bibitem{Dotter2008}
\bibinfo{author}{{Dotter}, A.} \emph{et~al.}
\newblock \bibinfo{journal}{\bibinfo{title}{{The Dartmouth Stellar Evolution
  Database}}}.
\newblock {\emph{\JournalTitle{\apjs}}} \textbf{\bibinfo{volume}{178}},
  \bibinfo{pages}{89--101}, \doiprefix\url{10.1086/589654}
  (\bibinfo{year}{2008}).
\newblock \eprint{0804.4473}.

\bibitem{Feiden2011}
\bibinfo{author}{{Feiden}, G.~A.}, \bibinfo{author}{{Chaboyer}, B.} \&
  \bibinfo{author}{{Dotter}, A.}
\newblock \bibinfo{journal}{\bibinfo{title}{{Accurate Low-mass Stellar Models
  of KOI-126}}}.
\newblock {\emph{\JournalTitle{\apjl}}} \textbf{\bibinfo{volume}{740}},
  \bibinfo{pages}{L25}, \doiprefix\url{10.1088/2041-8205/740/1/L25}
  (\bibinfo{year}{2011}).
\newblock \eprint{1109.2063}.

\bibitem{Baraffe1997}
\bibinfo{author}{{Baraffe}, I.}, \bibinfo{author}{{Chabrier}, G.},
  \bibinfo{author}{{Allard}, F.} \& \bibinfo{author}{{Hauschildt}, P.~H.}
\newblock \bibinfo{journal}{\bibinfo{title}{{Evolutionary models for metal-poor
  low-mass stars. Lower main sequence of globular clusters and halo field
  stars}}}.
\newblock {\emph{\JournalTitle{\aap}}} \textbf{\bibinfo{volume}{327}},
  \bibinfo{pages}{1054--1069} (\bibinfo{year}{1997}).
\newblock \eprint{astro-ph/9704144}.

\end{thebibliography}

\begin{thebibliography}{10}
\urlstyle{rm}
\expandafter\ifx\csname url\endcsname\relax
  \def\url#1{\texttt{#1}}\fi
\expandafter\ifx\csname urlprefix\endcsname\relax\def\urlprefix{URL }\fi
\expandafter\ifx\csname doiprefix\endcsname\relax\def\doiprefix{DOI: }\fi
\providecommand{\bibinfo}[2]{#2}
\providecommand{\eprint}[2][]{\url{#2}}

\bibitem{vernet2011}
\bibinfo{author}{{Vernet}, J.} \emph{et~al.}
\newblock \bibinfo{journal}{\bibinfo{title}{{X-shooter, the new wide band
  intermediate resolution spectrograph at the ESO Very Large Telescope}}}.
\newblock {\emph{\JournalTitle{\aap}}} \textbf{\bibinfo{volume}{536}},
  \bibinfo{pages}{A105}, \doiprefix\url{10.1051/0004-6361/201117752}
  (\bibinfo{year}{2011}).
\newblock \eprint{1110.1944}.

\bibitem{dhillon2018}
\bibinfo{author}{{Dhillon}, V.} \emph{et~al.}
\newblock \bibinfo{title}{{First light with HiPERCAM on the GTC}}.
\newblock In \emph{\bibinfo{booktitle}{Society of Photo-Optical Instrumentation
  Engineers (SPIE) Conference Series}}, vol. \bibinfo{volume}{10702} of
  \emph{\bibinfo{series}{Society of Photo-Optical Instrumentation Engineers
  (SPIE) Conference Series}}, \bibinfo{pages}{107020L},
  \doiprefix\url{10.1117/12.2312041} (\bibinfo{year}{2018}).
\newblock \eprint{1807.00557}.

\bibitem{Rebassa2007}
\bibinfo{author}{{Rebassa-Mansergas}, A.}, \bibinfo{author}{{G{\"a}nsicke},
  B.~T.}, \bibinfo{author}{{Rodr{\'{\i}}guez-Gil}, P.},
  \bibinfo{author}{{Schreiber}, M.~R.} \& \bibinfo{author}{{Koester}, D.}
\newblock \bibinfo{journal}{\bibinfo{title}{{Post-common-envelope binaries from
  SDSS - I. 101 white dwarf main-sequence binaries with multiple Sloan Digital
  Sky Survey spectroscopy}}}.
\newblock {\emph{\JournalTitle{\mnras}}} \textbf{\bibinfo{volume}{382}},
  \bibinfo{pages}{1377--1393}, \doiprefix\url{10.1111/j.1365-2966.2007.12288.x}
  (\bibinfo{year}{2007}).
\newblock \eprint{0707.4107}.

\bibitem{rechenberg1994}
\bibinfo{author}{{Rechenberg}, I.}
\newblock \emph{\bibinfo{title}{Evolutionsstrategie '94}}
  (\bibinfo{publisher}{Froomann--Holzboog}, \bibinfo{address}{Stuttgart},
  \bibinfo{year}{1994}).

\bibitem{Rajpurohit2016}
\bibinfo{author}{{Rajpurohit}, A.~S.} \emph{et~al.}
\newblock \bibinfo{journal}{\bibinfo{title}{{Spectral energy distribution of
  M-subdwarfs: A study of their atmospheric properties}}}.
\newblock {\emph{\JournalTitle{\aap}}} \textbf{\bibinfo{volume}{596}},
  \bibinfo{pages}{A33}, \doiprefix\url{10.1051/0004-6361/201526776}
  (\bibinfo{year}{2016}).
\newblock \eprint{1609.07062}.

\bibitem{Kesseli2018}
\bibinfo{author}{{Kesseli}, A.~Y.} \emph{et~al.}
\newblock \bibinfo{journal}{\bibinfo{title}{{Radii of 88 M Subdwarfs and
  Updated Radius Relations for Low-Metallicity M Dwarf Stars}}}.
\newblock {\emph{\JournalTitle{ArXiv e-prints}}}  (\bibinfo{year}{2018}).
\newblock \eprint{1810.07702}.

\bibitem{Koester2010}
\bibinfo{author}{{Koester}, D.}
\newblock \bibinfo{journal}{\bibinfo{title}{White dwarf spectra and atmosphere
  models}}.
\newblock {\emph{\JournalTitle{Memorie della Societa Astronomica Italiana}}}
  \textbf{\bibinfo{volume}{81}}, \bibinfo{pages}{921--931}
  (\bibinfo{year}{2010}).

\bibitem{Rebassa2017}
\bibinfo{author}{{Rebassa-Mansergas}, A.} \emph{et~al.}
\newblock \bibinfo{journal}{\bibinfo{title}{{The white dwarf binary pathways
  survey - II. Radial velocities of 1453 FGK stars with white dwarf companions
  from LAMOST DR 4}}}.
\newblock {\emph{\JournalTitle{\mnras}}} \textbf{\bibinfo{volume}{472}},
  \bibinfo{pages}{4193--4203}, \doiprefix\url{10.1093/mnras/stx2259}
  (\bibinfo{year}{2017}).
\newblock \eprint{1708.09480}.

\bibitem{Scargle1982}
\bibinfo{author}{{Scargle}, J.~D.}
\newblock \bibinfo{journal}{\bibinfo{title}{Studies in astronomical time series
  analysis. ii~- statistical aspects of spectral analysis of unevenly spaced
  data}}.
\newblock {\emph{\JournalTitle{Astrophys. J.}}} \textbf{\bibinfo{volume}{263}},
  \bibinfo{pages}{835--853} (\bibinfo{year}{1982}).

\bibitem{Schwarzenberg-czerny1996}
\bibinfo{author}{{Schwarzenberg-Czerny}, A.}
\newblock \bibinfo{journal}{\bibinfo{title}{Fast and statistically optimal
  period search in uneven sampled observations}}.
\newblock {\emph{\JournalTitle{\apjl}}} \textbf{\bibinfo{volume}{460}},
  \bibinfo{pages}{L107--L110} (\bibinfo{year}{1996}).

\bibitem{copperwheat2010}
\bibinfo{author}{{Copperwheat}, C.~M.} \emph{et~al.}
\newblock \bibinfo{journal}{\bibinfo{title}{{Physical properties of IP Pegasi:
  an eclipsing dwarf nova with an unusually cool white dwarf}}}.
\newblock {\emph{\JournalTitle{\mnras}}} \textbf{\bibinfo{volume}{402}},
  \bibinfo{pages}{1824--1840}, \doiprefix\url{10.1111/j.1365-2966.2009.16010.x}
  (\bibinfo{year}{2010}).
\newblock \eprint{0911.1637}.

\bibitem{Bloemen2011}
\bibinfo{author}{{Bloemen}, S.} \emph{et~al.}
\newblock \bibinfo{journal}{\bibinfo{title}{{Kepler observations of the beaming
  binary KPD 1946+4340}}}.
\newblock {\emph{\JournalTitle{\mnras}}} \textbf{\bibinfo{volume}{410}},
  \bibinfo{pages}{1787--1796}, \doiprefix\url{10.1111/j.1365-2966.2010.17559.x}
  (\bibinfo{year}{2011}).
\newblock \eprint{1010.2747}.

\bibitem{claret2000}
\bibinfo{author}{{Claret}, A.}
\newblock \bibinfo{journal}{\bibinfo{title}{{Non-linear limb-darkening law for
  LTE models (Claret, 2000)}}}.
\newblock {\emph{\JournalTitle{VizieR Online Data Catalog}}}
  \textbf{\bibinfo{volume}{336}}, \bibinfo{pages}{31081}
  (\bibinfo{year}{2000}).

\bibitem{gianninas2013}
\bibinfo{author}{{Gianninas}, A.}, \bibinfo{author}{{Strickland}, B.~D.},
  \bibinfo{author}{{Kilic}, M.} \& \bibinfo{author}{{Bergeron}, P.}
\newblock \bibinfo{journal}{\bibinfo{title}{{Limb-darkening Coefficients for
  Eclipsing White Dwarfs}}}.
\newblock {\emph{\JournalTitle{\apj}}} \textbf{\bibinfo{volume}{766}},
  \bibinfo{pages}{3}, \doiprefix\url{10.1088/0004-637X/766/1/3}
  (\bibinfo{year}{2013}).
\newblock \eprint{1301.7091}.

\bibitem{claret2011}
\bibinfo{author}{{Claret}, A.} \& \bibinfo{author}{{Bloemen}, S.}
\newblock \bibinfo{journal}{\bibinfo{title}{{Gravity and limb-darkening
  coefficients for the Kepler, CoRoT, Spitzer, uvby, UBVRIJHK, and Sloan
  photometric systems}}}.
\newblock {\emph{\JournalTitle{\aap}}} \textbf{\bibinfo{volume}{529}},
  \bibinfo{pages}{A75}, \doiprefix\url{10.1051/0004-6361/201116451}
  (\bibinfo{year}{2011}).

\bibitem{parsons2017}
\bibinfo{author}{{Parsons}, S.~G.} \emph{et~al.}
\newblock \bibinfo{journal}{\bibinfo{title}{{Testing the white dwarf
  mass-radius relationship with eclipsing binaries}}}.
\newblock {\emph{\JournalTitle{\mnras}}} \textbf{\bibinfo{volume}{470}},
  \bibinfo{pages}{4473--4492}, \doiprefix\url{10.1093/mnras/stx1522}
  (\bibinfo{year}{2017}).
\newblock \eprint{1706.05016}.

\bibitem{panei2007}
\bibinfo{author}{{Panei}, J.~A.}, \bibinfo{author}{{Althaus}, L.~G.},
  \bibinfo{author}{{Chen}, X.} \& \bibinfo{author}{{Han}, Z.}
\newblock \bibinfo{journal}{\bibinfo{title}{{Full evolution of low-mass white
  dwarfs with helium and oxygen cores}}}.
\newblock {\emph{\JournalTitle{\mnras}}} \textbf{\bibinfo{volume}{382}},
  \bibinfo{pages}{779--792}, \doiprefix\url{10.1111/j.1365-2966.2007.12400.x}
  (\bibinfo{year}{2007}).

\bibitem{Althaus2017}
\bibinfo{author}{{Althaus}, L.~G.}, \bibinfo{author}{{De Ger{\'o}nimo}, F.},
  \bibinfo{author}{{C{\'o}rsico}, A.}, \bibinfo{author}{{Torres}, S.} \&
  \bibinfo{author}{{Garc{\'{\i}}a-Berro}, E.}
\newblock \bibinfo{journal}{\bibinfo{title}{{The evolution of white dwarfs
  resulting from helium-enhanced, low-metallicity progenitor stars}}}.
\newblock {\emph{\JournalTitle{\aap}}} \textbf{\bibinfo{volume}{597}},
  \bibinfo{pages}{A67}, \doiprefix\url{10.1051/0004-6361/201629909}
  (\bibinfo{year}{2017}).
\newblock \eprint{1611.06191}.

\bibitem{press2007}
\bibinfo{author}{{Press}, W.~H.}, \bibinfo{author}{{Teukolsky}, A.~A.},
  \bibinfo{author}{{Vetterling}, W.~T.} \& \bibinfo{author}{{Flannery}, B.~P.}
\newblock \emph{\bibinfo{title}{{Numerical recipes. The art of scientific
  computing, 3rd edn.}}} (\bibinfo{publisher}{Cambridge: University Press},
  \bibinfo{year}{2007}).

\bibitem{Parsons2018Md}
\bibinfo{author}{{Parsons}, S.~G.} \emph{et~al.}
\newblock \bibinfo{journal}{\bibinfo{title}{{The scatter of the M dwarf
  mass-radius relationship}}}.
\newblock {\emph{\JournalTitle{\mnras}}} \textbf{\bibinfo{volume}{481}},
  \bibinfo{pages}{1083--1096}, \doiprefix\url{10.1093/mnras/sty2345}
  (\bibinfo{year}{2018}).
\newblock \eprint{1808.07780}.

\bibitem{Allard2012}
\bibinfo{author}{{Allard}, F.}, \bibinfo{author}{{Homeier}, D.} \&
  \bibinfo{author}{{Freytag}, B.}
\newblock \bibinfo{journal}{\bibinfo{title}{{Models of very-low-mass stars,
  brown dwarfs and exoplanets}}}.
\newblock {\emph{\JournalTitle{Philosophical Transactions of the Royal Society
  of London Series A}}} \textbf{\bibinfo{volume}{370}},
  \bibinfo{pages}{2765--2777}, \doiprefix\url{10.1098/rsta.2011.0269}
  (\bibinfo{year}{2012}).
\newblock \eprint{1112.3591}.

\bibitem{Schlafly+Finkbeiner11}
\bibinfo{author}{{Schlafly}, E.~F.} \& \bibinfo{author}{{Finkbeiner}, D.~P.}
\newblock \bibinfo{journal}{\bibinfo{title}{{Measuring Reddening with Sloan
  Digital Sky Survey Stellar Spectra and Recalibrating SFD}}}.
\newblock {\emph{\JournalTitle{\apj}}} \textbf{\bibinfo{volume}{737}},
  \bibinfo{pages}{103}, \doiprefix\url{10.1088/0004-637X/737/2/103}
  (\bibinfo{year}{2011}).
\newblock \eprint{1012.4804}.

\bibitem{Bailer2018}
\bibinfo{author}{{Bailer-Jones}, C.~A.~L.}, \bibinfo{author}{{Rybizki}, J.},
  \bibinfo{author}{{Fouesneau}, M.}, \bibinfo{author}{{Mantelet}, G.} \&
  \bibinfo{author}{{Andrae}, R.}
\newblock \bibinfo{journal}{\bibinfo{title}{{Estimating Distance from
  Parallaxes. IV. Distances to 1.33 Billion Stars in Gaia Data Release 2}}}.
\newblock {\emph{\JournalTitle{\aj}}} \textbf{\bibinfo{volume}{156}},
  \bibinfo{pages}{58}, \doiprefix\url{10.3847/1538-3881/aacb21}
  (\bibinfo{year}{2018}).
\newblock \eprint{1804.10121}.

\bibitem{Gaia2018}
\bibinfo{author}{{Gaia Collaboration}} \emph{et~al.}
\newblock \bibinfo{journal}{\bibinfo{title}{{Gaia Data Release 2. Summary of
  the contents and survey properties}}}.
\newblock {\emph{\JournalTitle{\aap}}} \textbf{\bibinfo{volume}{616}},
  \bibinfo{pages}{A1}, \doiprefix\url{10.1051/0004-6361/201833051}
  (\bibinfo{year}{2018}).
\newblock \eprint{1804.09365}.

\bibitem{Blanco2014}
\bibinfo{author}{{Blanco-Cuaresma}, S.}, \bibinfo{author}{{Soubiran}, C.},
  \bibinfo{author}{{Heiter}, U.} \& \bibinfo{author}{{Jofr{\'e}}, P.}
\newblock \bibinfo{journal}{\bibinfo{title}{{Determining stellar atmospheric
  parameters and chemical abundances of FGK stars with iSpec}}}.
\newblock {\emph{\JournalTitle{\aap}}} \textbf{\bibinfo{volume}{569}},
  \bibinfo{pages}{A111}, \doiprefix\url{10.1051/0004-6361/201423945}
  (\bibinfo{year}{2014}).
\newblock \eprint{1407.2608}.

\bibitem{Gustafsson2008}
\bibinfo{author}{{Gustafsson}, B.} \emph{et~al.}
\newblock \bibinfo{journal}{\bibinfo{title}{{A grid of MARCS model atmospheres
  for late-type stars. I. Methods and general properties}}}.
\newblock {\emph{\JournalTitle{\aap}}} \textbf{\bibinfo{volume}{486}},
  \bibinfo{pages}{951--970}, \doiprefix\url{10.1051/0004-6361:200809724}
  (\bibinfo{year}{2008}).
\newblock \eprint{0805.0554}.

\bibitem{Izzard2004}
\bibinfo{author}{{Izzard}, R.~G.}, \bibinfo{author}{{Tout}, C.~A.},
  \bibinfo{author}{{Karakas}, A.~I.} \& \bibinfo{author}{{Pols}, O.~R.}
\newblock \bibinfo{journal}{\bibinfo{title}{{A new synthetic model for
  asymptotic giant branch stars}}}.
\newblock {\emph{\JournalTitle{\mnras}}} \textbf{\bibinfo{volume}{350}},
  \bibinfo{pages}{407--426}, \doiprefix\url{10.1111/j.1365-2966.2004.07446.x}
  (\bibinfo{year}{2004}).
\newblock \eprint{astro-ph/0402403}.

\bibitem{Izzard2018}
\bibinfo{author}{{Izzard}, R.~G.} \emph{et~al.}
\newblock \bibinfo{journal}{\bibinfo{title}{{Binary stars in the Galactic thick
  disc}}}.
\newblock {\emph{\JournalTitle{\mnras}}} \textbf{\bibinfo{volume}{473}},
  \bibinfo{pages}{2984--2999}, \doiprefix\url{10.1093/mnras/stx2355}
  (\bibinfo{year}{2018}).
\newblock \eprint{1709.05237}.

\bibitem{Hurley2002}
\bibinfo{author}{{Hurley}, J.~R.}, \bibinfo{author}{{Tout}, C.~A.} \&
  \bibinfo{author}{{Pols}, O.~R.}
\newblock \bibinfo{journal}{\bibinfo{title}{{Evolution of binary stars and the
  effect of tides on binary populations}}}.
\newblock {\emph{\JournalTitle{\mnras}}} \textbf{\bibinfo{volume}{329}},
  \bibinfo{pages}{897--928}, \doiprefix\url{10.1046/j.1365-8711.2002.05038.x}
  (\bibinfo{year}{2002}).
\newblock \eprint{astro-ph/0201220}.

\end{thebibliography}

\section*{Acknowledgements}

This  work was  partially  supported  by the  MINECO  Ram\'on y  Cajal
programme RYJ-2016-20254 (A.R.M.) and grant AYA\-2017-86274-P (A.R.M.,
S.T.)   and   by   the   the   AGAUR   grant   SGR-661/2017   (A.R.M.,
S.T.).   S.G.P.   acknowledges   the   support   of   the   Leverhulme
Trust.  J.J.R.  acknowledges  support  from the  Joint  Funds  of  the
National Natural Sciences Foundation of  China (Grant Nos U1531244 and
U1831209), the NSFC  grant 11833006 and the Young  Researcher Grant of
National    Astronomical    Observatories,    Chinese    Academy    of
Sciences.  HiPERCAM and  V.S.D  are funded  by  the European  Research
Council  under  the  European   Union’s  Seventh  Framework  Programme
(FP/2007-2013)   under  ERC-2013-ADG   Grant  Agreement   no.   340040
(HiPERCAM).

The authors thank Fran Jim\'emez-Esteban for valuable input in the use
of  VOSA, Aurora  Kesseli  for  sharing her  usd  spectra and  Leandro
Althaus for helpful discussions.

\section*{Author contributions statement}

All  authors   have  contributed  to   the  work  presented   in  this
paper. A.R.M. performed the decomposition  and fitting of the spectra,
carried out the  entire spectral analysis (except the  one required by
iSpec),  conducted  the VOSA  analysis  and  led  the writing  of  the
manuscript. S.G.P. reduced all  the spectroscopic and photometric data
and carried out the light-curve  analysis. V.S.D. and S.P.L. performed
the  GTC observations.  J.J.R. conducted  the iSpec  analysis. V.S.D.,
S.P.L. and T.R.M. contributed to  the development of HiPERCAM, a vital
instrument for  obtaining the results  of this work. S.T.  carried out
the binary$\_c$ simulation and calculated the cooling age of the white
dwarf. A.R.M.,  S.G.P. and J.J.R.  discovered the system.  All authors
reviewed the manuscript.

\section*{Additional information}

\textbf{Competing Interests}.  The authors  declare that they  have no
competing financial interests.

\begin{figure}[ht]
\centering
\includegraphics[width=0.7\linewidth]{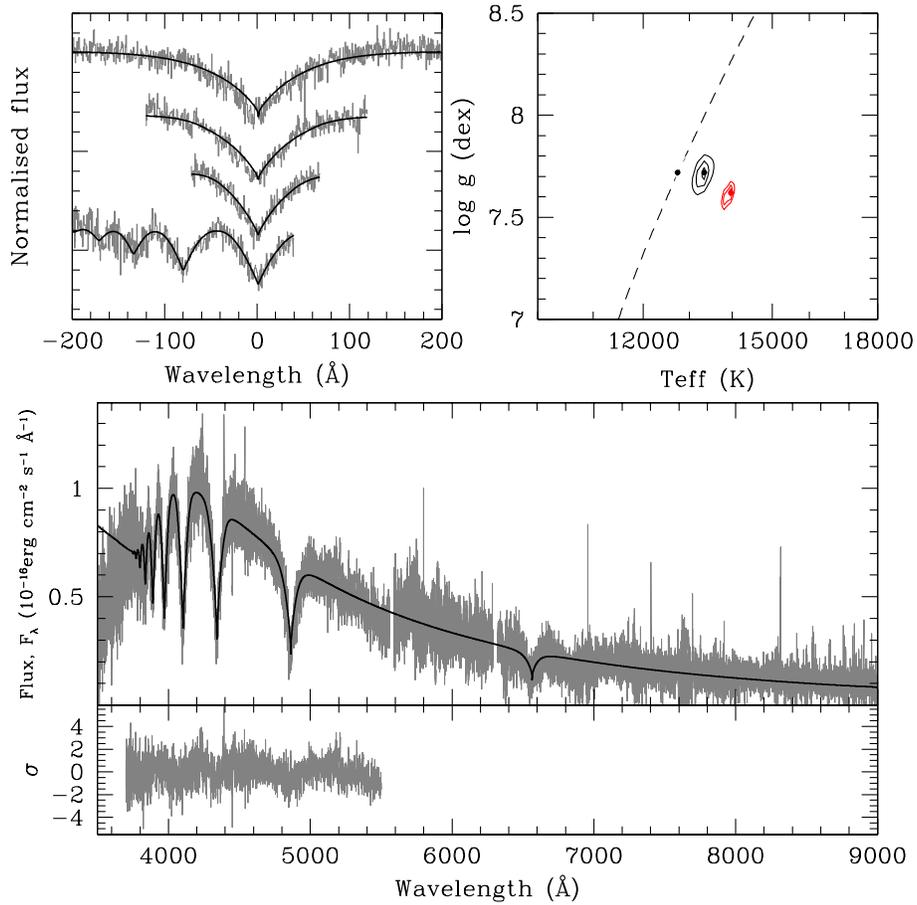}
\caption{Spectral model fits  to the residual white  dwarf spectrum of
  SDSS\,J2355+0448,  obtained  after  subtracting  the  best-fit  cool
  subdwarf template (Figure\,1).   Top left panel: from  top to bottom
  the best-fit (black  line) to the normalised H$\beta$  to H10 (gray)
  line  profiles.  Top  right  panel: 3,  5,  and 10$\sigma$  $\chi^2$
  contour plots in the  T$_\mathrm{eff}-\log g$ plane.  Black contours
  indicate  the  best  line  profile fits,  whilst  red  contours  the
  best-fit to  the entire  spectrum.  The maximum  H$\beta$ equivalent
  width is indicated  by a dashed line.  Black dots  indicate the best
  ``hot'' and ``cold''  line profile solutions, the  red dot indicates
  the  best fit  to the  whole spectrum.   Bottom panel:  the residual
  white dwarf spectrum after subtracting the cool subdwarf (gray line)
  together with  the best-fit white  dwarf model (black line)  and the
  residuals  (gray  line, bottom).   The  fit  to the  whole  spectrum
  selects the ``hot'' solution.}
\label{fig:fitwd}
\end{figure}

\begin{figure}[ht]
\centering
\includegraphics[angle=-90,width=0.6\linewidth]{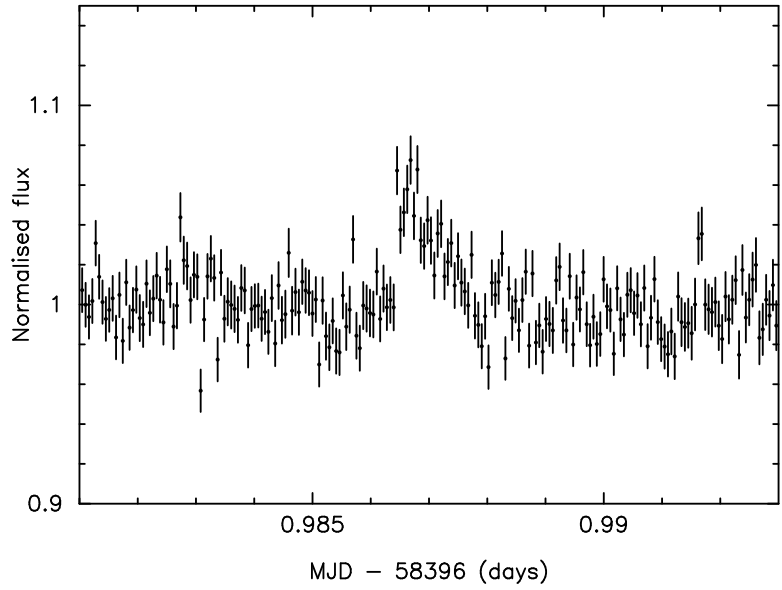}
\caption{$g_s$-band    HiPERCAM   light-curve    of   SDSS\,J2355+0448
  displaying a flare  originating in the surface of  the cool subdwarf
  star. The error bars represent $\pm1\sigma$ uncertainties.}
\label{fig:flare}
\end{figure}

\begin{figure}[ht]
\centering
\includegraphics[angle=-90,width=0.7\linewidth]{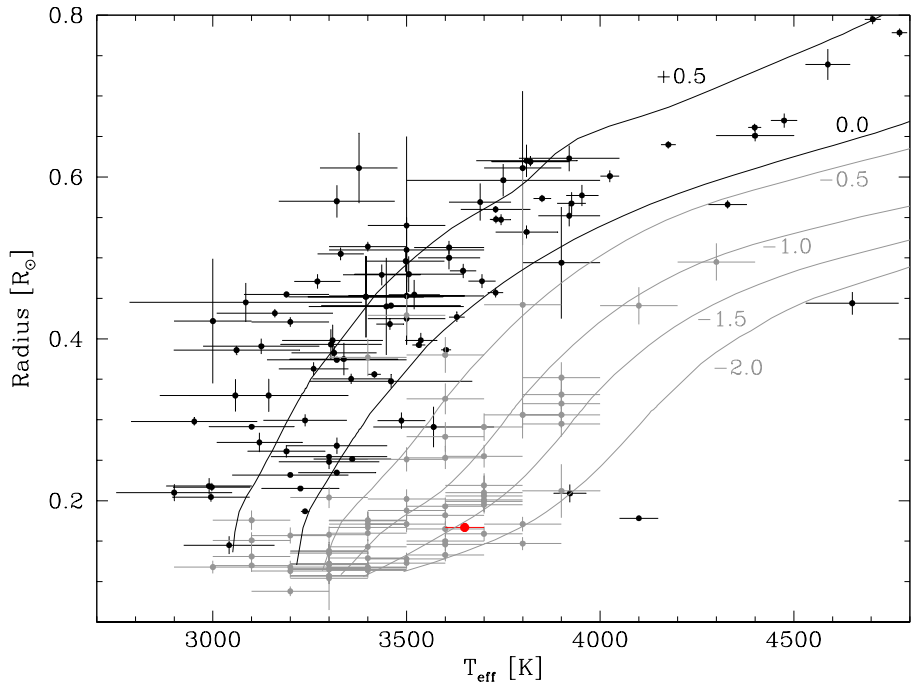}
\caption{Radius-effective  temperature  plot   for  a  compilation  of
  low-mass solar-metallicity  stars (black)  and for a  compilation of
  cool subdwarfs (grey). The solid lines are the Dartmouth theoretical
  tracks  for 10  Gyr  and  the indicated  [Fe/H]  abundances (in  dex
  units).  The red  solid dot represents the cool  subdwarf studied in
  this work. The error bars represent $\pm1\sigma$ uncertainties.}
\label{fig:t-r}
\end{figure}

\begin{figure}[ht]
\centering
\includegraphics[angle=-90,width=0.6\linewidth]{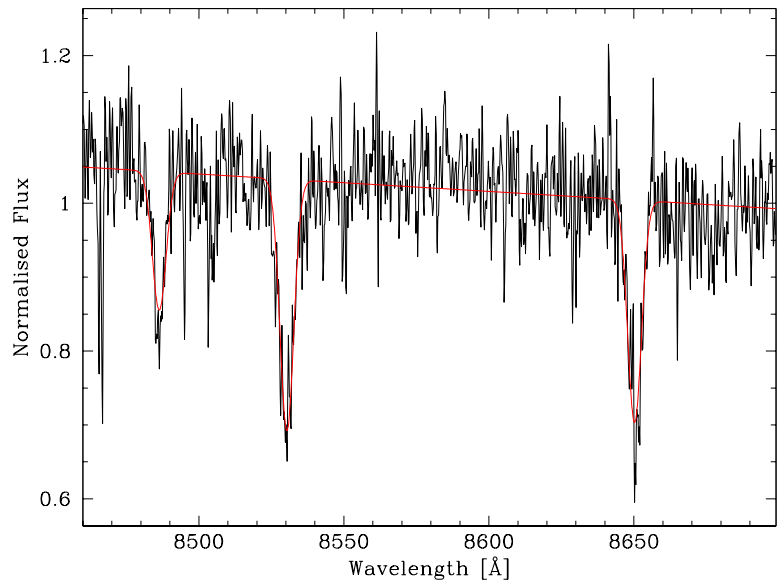}
\caption{Fit (red solid line) to the normalised \Ion{Ca}{II} triplet
  at $\sim$8500\AA\ of a X-Shooter spectrum (black solid lines).}
\label{fig:rvfit}
\end{figure}

\begin{table}[ht]
\centering
\begin{tabular}{@{}lcccc@{}}
\hline
Parameter & $g_s$ & $r_s$ & $i_s$ & $z_s$  \\
\hline
WD $a_1$ & 0.6882 & 0.6126 & 0.5782 & 0.5530 \\
WD $a_2$ & 0.1635 & 0.0232 & -0.1761 & -0.3300 \\
WD $a_3$ & -0.4165 & -0.1818 & 0.0999 & 0.3024 \\
WD $a_4$ & 0.1915 & 0.0823 & -0.0386 & -0.1209 \\
sd $a_1$ & -0.4633 & -0.6869 & -0.3228 & -0.1407 \\
sd $a_2$ & 2.3583 & 3.9532 & 2.9078 & 2.2760 \\
sd $a_3$ & -1.4110 & -3.7931 & -2.8289 & -2.2345 \\
sd $a_4$ & 0.1903 & 1.1713 & 0.8566 & 0.6676 \\
sd grav. darkening & 1.1211 & 0.7539 & 0.6513 & 0.5731 \\
\hline
\end{tabular}
\caption{\label{tab:darkening_coefs}   Limb-   and   gravity-darkening
  coefficients  used during  our light  curve fitting  for the  $g_s$,
  $r_s$,  $i_s$  and  $z_s$  bands.  The  white  dwarf  limb-darkening
  coefficients (WD  $a_1$ to  WD $a_4$)  are for  a $\teff=13,250$\,K,
  $\logg=7.75$ dex  white dwarf. The cool  subdwarf limb-darkening (sd
  $a_1$  to  sd  $a_4$)  and surface  gravity  (sd  grav.   darkening)
  coefficients   are   for   a  $\teff=3,650$\,K,   $\logg=5.0$   dex,
  [Fe/H]$=-2.0$ dex star.}
\end{table}

\end{document}